\documentclass[a4paper,usenames,dvipsnames,11pt]{article}
\pdfoutput=1
\usepackage [latin1]{inputenc}
\usepackage{jheppub}
\usepackage{slashed}
\usepackage{mathrsfs,booktabs,multirow,tabularx}
\usepackage{stmaryrd}
\usepackage{xspace}
\usepackage{fancyvrb}
\usepackage[makeroom]{cancel}
\usepackage{amsmath}    % need for subequations
\usepackage{amssymb}    % 
\usepackage{graphicx}   % need for figures
\usepackage{verbatim}   % useful for program listings
\usepackage{lscape}
\usepackage{subfig}
\usepackage{listings}
\usepackage{hyperref}

\def\beq{\begin{equation}}
\def\eeq{\end{equation}}
\def\beqn{\begin{eqnarray}}
\def\eeqn{\end{eqnarray}}

\def\NEPbw#1{{\rm NEP}_{#1}}
\def\NEPbwR#1{{\rm NEP}_{#1}^{({\rm R})}}
\def\NEPbwE#1{{\rm NEP}_{#1}^{({\rm E})}}

\newcommand\sss{\scriptscriptstyle}

\newcommand\ep{\epsilon}
\newcommand\half{\frac{1}{2}}

\newcommand\as{\alpha_{\sss S}}
\newcommand\asotpi{\frac{\as}{2\pi}}
\newcommand\tpioas{\frac{2\pi}{\as}}

\newcommand\aem{\alpha}

\newcommand\ord{{\cal O}}
\newcommand{\Oop}{{\mathbb O}}
\newcommand{\mll}{M_{ll}}
\newcommand{\mut}{\mu^2}
\newcommand{\muz}{\mu_0}
\newcommand{\muzt}{\muz^2}
\newcommand{\hP}{\hat{P}}

\newcommand\NF{N_{\sss F}}
\newcommand\CF{C_{\sss F}}

\newcommand\CA{C_{\sss A}}
\newcommand\TF{T_{\sss F}}

\newcommand\Nu{N_u}
\newcommand\Nd{N_d}
\newcommand\MSb{\overline{\rm MS}}
\newcommand{\Aop}{{\mathbb A}}
\newcommand{\Bop}{{\mathbb B}}
\newcommand{\Cop}{{\mathbb C}}
\newcommand{\Wop}{{\mathbb W}}
\newcommand{\Zop}{{\mathbb Z}}
\newcommand{\Sop}{{\mathbb S}}

\newcommand{\Top}{{\mathbb T}}
\newcommand\APmat{{\mathbb P}}
\newcommand\stepf{\Theta}
\newcommand\bq{\bar{q}}
\newcommand\hsig{\hat{\sigma}}
\newcommand\epLij{\epsilon_{ij}^{{\sss\rm L}}}
\newcommand\epUij{\epsilon_{ij}^{{\sss\rm U}}}
\newcommand{\qz}{q_0}
\newcommand{\qzt}{\qz^2}
\newcommand{\NEPbwi}{{\rm NEP}_i}
\newcommand{\NEPbwiR}{{\rm NEP}_i^{({\rm R})}}
\newcommand{\NEPbwiE}{{\rm NEP}_i^{({\rm E})}}

\newcommand\Rin{{\sss\rm IN}}
\newcommand\Rout{{\sss\rm OUT}}
\newcommand{\Rinout}{\mathrel{\vcenter{\offinterlineskip
 \ialign{##\cr${\sss\rm ~IN}$\cr\noalign{\kern+0.5pt}${\sss\rm OUT}$\cr}}}}

\title{Correcting for cutoff dependence in backward evolution of QCD parton
showers}

\author[a]{S. Frixione,}
\author[b]{B.R. Webber}
\affiliation[a]{INFN, Sezione di Genova, Via Dodecaneso 33, I-16146, 
Genoa, Italy}
\affiliation[b]{University of Cambridge, Cavendish Laboratory,
  J.J.~Thomson Avenue, Cambridge, UK}

\emailAdd{Stefano.Frixione@cern.ch}
\emailAdd{webber@hep.phy.cam.ac.uk}

\abstract{
Monte Carlo event generators for hard hadronic collisions depend
on the evolution of parton showers backwards from a high-scale subprocess to
the hadronization scale.  The evolution is treated as a branching process
with a sequence of resolvable parton emissions. The criterion of
resolvability involves cutoffs that determine the no-emission probability
(NEP) for a given range of the evolution scale. Existing event generators
neglect cutoff-dependent terms in the NEP that, although formally
power-suppressed, can have significant phenomenological effects.  We compute
such terms and study their consequences. One important result is that it 
is not possible for the backward shower to faithfully reproduce the 
cutoff-independent parton distribution functions (PDFs) used to generate it. 
We show that the computed NEP corrections mitigate but do not eliminate this 
problem. An alternative approach is to use cutoff-dependent PDFs that are 
consistent with the uncorrected NEP. Then one must apply cutoff-dependent 
corrections to hard subprocess matrix elements. We compute those corrections 
to the first nontrivial order for the Drell-Yan process and for Higgs 
production by gluon fusion.
}


\begin{document}
\maketitle
\flushbottom

\section{Introduction\label{sec:intro}}
Although the precision of predictions of short-distance cross sections
using QCD perturbation theory has greatly increased in recent
years, it remains true that their comparison with experimental data
relies to a large extent on parton-shower based Monte Carlo event
generators\footnote{For a  review see~\cite{Buckley:2019kjt}.} (MCEGs)
for the estimation of non-perturbative and approximate
higher-order perturbative effects. The connection between a measured 
cross section at a hadron collider and that stemming from the short-distance
subprocess can be described in an inclusive sense using perturbation
theory, factorization theorems and parton
distribution functions (PDFs).  However, for the more exclusive
description required for the estimation of experimental effects a
reliable MCEG is essential.

A key component of any MCEG for hadronic collisions is a
{\em backward parton shower}~\cite{Sjostrand:1985xi}
that links the short-distance subprocess to the incoming
hadrons via an iterative parton branching procedure.  For reasons of
Monte Carlo efficiency, the shower starts at the high virtuality scale
of the subprocess, with appropriate parton flavours and momentum
fractions, and ends at the lower scale of hadron formation. For
example, in the production of a $Z^0$ boson at leading order, the
parton showers should be initiated by a quark-antiquark pair of equal
flavour with invariant mass within the $Z^0$ line width.  If the showers were
generated forwards from the hadron scale, as is normally done in PDF
evolution, then the efficiency for finding a pair of the same flavour
with an appropriate invariant mass would be unacceptably low.

A special feature of the backward parton shower~\cite{Sjostrand:1985xi,
Marchesini:1987cf} is that it must be ``guided'' by input PDFs, which are 
supposed to ensure that the ensemble of parton flavours and momentum fractions 
in the shower at any intermediate scale remains consistent with 
those PDFs. Compared to forward evolution, this implies modifications
to both the probability of branching as a function of scale, and the
distribution of momentum fractions within each branching; these
modifications are not an option designed to improve efficiency, but
rather an inevitable feature of backward showering.  However, the
branching process necessarily involves a sequence of {\em resolvable}
parton emissions, defined by some cutoffs, while the
PDFs are normally taken from global fits that satisfy evolution
equations~\cite{Gribov:1972ri,Altarelli:1977zs,Dokshitzer:1977sg} that
contain no such cutoffs. This could give rise to systematic biases that,
as far as we are aware, have not been studied so far and are the focus of
the present paper.

In sect.~\ref{sec:gen} we present a general analysis of PDF evolution
equations, not limited to QCD or any particular perturbative order but
suited to the discussion of issues related to the resolvability of
emissions. We pay particular attention to the ambiguities in the
treatment of unresolved and virtual contributions, and the choices
inherent in their resolution.  Section~\ref{sec:MCbw} examines the
backward MC showering process in this framework, in particular the
key concept of the non-emission probability (NEP), which governs
the evolution of the shower in a way supposedly consistent with a
given set of guiding PDFs.  We show that neither of the NEP
expressions in current use is formally correct in the presence of
cutoffs. However, we find that there is no fully satisfactory
formulation of the NEP as long as the guiding PDFs satisfy the normal
cutoff-independent evolution equations.

In the generic shower, any splitting is achieved by means of the 
generation of three variables, which can always be mapped to fit the
following description (stemming mainly from a $1\to 2$ branching
example). One of the variables controls the relative angle between
the plane where the emission takes place and a given reference plane;
it does not necessitate any cutoff, and as far as PDF evolution is
concerned it is irrelevant. A second variable has canonical dimension
equal to one, and is generally identified as ``the'' shower evolution 
variable; it coincides with, or is strictly related to, the variable 
with respect to which the
derivative of the PDFs is evaluated in the PDF evolution equations.
In MCEGs it is constrained to be larger than a given scale of the order
of the typical hadron mass. Such a constraint is imposed by means of
a cutoff; it is clear that below or around the cutoff the PDFs lose 
physical meaning, both as reconstructed by the MCEG or as obtained in the 
evolution: thus, the corresponding cutoff will not concerns us here.

The third variable in the shower has canonical dimension equal to zero;
it coincides with, or is strictly related to, the variable that defines
the convolution product on the r.h.s.~of the PDF evolution equations,
which we typically denote by $z$ and refer to as the momentum
fraction. In MCEGs, cutoffs on this variable are unavoidable in order
for the very definition of the Sudakov form factors to be meaningful,
whereas they are absent from the DGLAP evolution equations; these are
the cutoffs we shall be concerned with henceforth.

The cutoffs on the momentum-fraction variable give rise to power
corrections in the PDFs generated by backward showering, which are
absent from those that satisfy the DGLAP equations. We emphasise that
we are not concerned here with evaluating the ``true'' power
corrections to PDFs, or rather to observables derived from them, which
are anyway process dependent.  Our concern is with the internal
consistency, or otherwise, of MCEGs that use PDFs for guiding backward
evolution, which generates in-principle different PDFs.\footnote{This is
  in contrast to the generation of fragmentation  functions (FFs) by
  forward parton showering, which does not use guiding FFs and
  therefore raises no issues of consistency.}

Section~\ref{sec:LO} applies the general approach of the preceding
sections to the case relevant to the most widely-used MCEGs (before
any matching or merging), namely that of leading-order QCD.  
We show results on the NEP formulations in current
use, the improved expression derived in sect.~\ref{sec:MCbw}, and
their effects on MC backward evolution.  The general conclusion is
that, while the improved expression performs best, all formulations
fail to achieve satisfactory consistency with the guiding
cutoff-independent PDFs.

We therefore turn in sect.~\ref{sec:PDFcut} to an alternative
approach, in which the guiding PDFs obey evolution equations that
incorporate the same cutoffs as the backward parton shower.  We show
that such PDFs can be made exactly consistent with the constraints of
flavour and momentum conservation, and verify that the corresponding
NEP ensures consistency between the guiding PDFs and the MC results.
Of course, if this approach were implemented, the cutoff-dependent
PDFs would be specific to the cutoffs employed in a particular MCEG,
and would need to be extracted from dedicated global fits.
Furthermore, in those fits the factorization of PDFs and
short-distance cross sections implies that the latter will also be
modified by cutoff-dependent terms.\footnote{This is also pointed out
  in ref.~\cite{Mendizabal:2023mel}.} In sect.~\ref{sec:xsecs} we
derive a general expression for these cutoff corrections to the
first nontrivial order in QCD, and illustrate its application to the
processes of lepton pair and Higgs boson production.  Finally in
sect.~\ref{sec:conc} we summarize our main results and conclusions.

Appendix~\ref{sec:PDFsol} contains a more detailed discussion  of the
relation between the evolution equations and the backward MC process.
Appendix~\ref{sec:purevirt} presents a toy model in which all emissions
are unresolvable, designed to illuminate the ambiguities and
difficulties in defining the NEP.

\section{PDF evolution equations\label{sec:gen}}
In this section, we recast the evolution equations for the PDFs in a 
form which is suited to the parton-shower-like approach adopted in
MCEGs. The evolution variable $\mut$ has canonical dimension
of mass squared; its specific nature is not relevant here.

The starting point is the evolution
equations~\cite{Gribov:1972ri,Altarelli:1977zs,Dokshitzer:1977sg},
 which we write as follows:
\beq
\frac{\partial F(x)}{\partial\log\mut}=\Oop\otimes_x F\,,
\label{evolmu2}
\eeq
where
\beq
\Oop\otimes_x F = \int_0^1 \frac{dz}{z}\,\Oop(z)\,F(x/z)\,,
\label{convdef}
\eeq
with the understanding that $F(x/z) = 0$ for $z<x$.
We assume to be working in a \mbox{$d\equiv 1+2(\Nu+\Nd)$}-dimensional
flavour space, where $F$ is a column vector whose $d$ individual components
$(F)_i$ are the PDFs $f_i$ of the various partons, and $\Oop$ is a $d\times d$ 
matrix, whose elements in the $\MSb$ factorisation scheme are the 
splitting kernels; additional terms are
present in a non-$\MSb$ factorisation scheme. Both $F$ and $\Oop$ 
are  $x$-space objects, that depend on $\mut$ as well; in the notation, 
either or both dependences may be included explicitly or understood. 
It is safe to assume, at least up to the NLO and in any factorisation
scheme, that the most general form of $\Oop$ is:
\beq
\Oop(z)=\left[\Aop(z)\right]_+ + \Bop\,\delta(1-z) + \Cop(z)\,,
\label{Oopform}
\eeq
where
\beq
\left(\Aop(z)\right)_{ij}=\delta_{ij}A_i(z)\,,\;\;\;\;\;\;\;\;
\left(\Bop\right)_{ij}=\delta_{ij}B_i\,,\;\;\;\;\;\;\;\;
\left(\Cop(z)\right)_{ij}=C_{ij}(z)\,,
\label{AopBopCop}
\eeq
with $1\le i,j\le d$ the parton indices; note that $\Cop$ is 
in general non-diagonal. $A_i(z)$ and $C_{ij}(z)$ are regular functions
of $z$, and $B_i$ are constants in $z$; all of them depend on $\mut$.
Typically, $A_i(z)$ diverges when $z\to 1$, and the plus prescription
in eq.~(\ref{Oopform}) regularises that divergence; any divergence
at $z\to 0$ is not regularised. Equation~(\ref{Oopform}) encompasses
one of the forms in which the NLO splitting kernels are usually 
written, namely:
\beq
\sum_{k=0}^1\left(\frac{\aem(\mut)}{2\pi}\right)^{k+1}\APmat^{[k]}(z)=
\widetilde{\Aop}(z)\left[\frac{1}{1-z}\right]_+ + 
\widetilde{\Bop}\,\delta(1-z) + \Cop(z)\,,
\label{Oopform2}
\eeq
with $\widetilde{\Aop}(z)$ finite at $z=1$. Indeed, it is a matter of 
applying the definition of the plus distribution 
to show that, {\em when} the following relationships
\beqn
\Aop(z)=\frac{\widetilde{\Aop}(z)}{1-z}\,,\;\;\;\;\;\;\;\;
\Bop=\widetilde{\Bop}+
\int_0^1 dz\,\frac{\widetilde{\Aop}(z)-\widetilde{\Aop}(1)}{1-z}
\label{ABopid}
\eeqn
hold, then the r.h.s.'s of eqs.~(\ref{Oopform}) and~(\ref{Oopform2}) 
are identical to one another. 

In order to proceed, we introduce the following symbols:
\beq
\stepf_{ij,z}^\Rin=\stepf(\epLij<z<1-\epUij)\,,\;\;\;\;\;\;\;\;
\stepf_{ij,z}^\Rout\equiv 1-\stepf_{ij,z}^\Rin=
\stepf(z<\epLij)+\stepf(z>1-\epUij)\,,
\label{thetapm}
\eeq
with $1-\epLij-\epUij>0$, $\epLij>0$, and $\epUij>0$. The parameters
$\epLij$ and $\epUij$ are flavour (and possibly scale) dependent cutoffs,
which help to define an inner ($\stepf_{ij,z}^\Rin$) and an outer 
($\stepf_{ij,z}^\Rout$) region; the former (latter) will be associated 
with resolved (unresolved) emissions in the $z$ space for the branching:
\beq
j(1)\;\longrightarrow\;i(z)+k(1-z)
\;\;\;\;\Longleftrightarrow\;\;\;\;
P_{ij}(z)\,.
\eeq
Equation~(\ref{thetapm}) then implies that $\epLij$ and $\epUij$
limit from below the fractional energy of parton $i$ and recoil
system $k$, respectively. At the LO, the recoil system is a parton
itself, unambiguously determined by $i$ and $j$, so that it may be
denoted by \mbox{$k=j\ominus i$}. This suggests introducing the 
\mbox{$1+\Nu+\Nd$} parameters:
\beq
\ep_g\,,\;\ep_u\,,\;\ep_d\,,\ldots\,,
\eeq
and setting:
\beq
\epLij=\ep_i\,,\;\;\;\;\;\;\;\;\epUij=\ep_{j\ominus i}\,.
\label{epflavLO}
\eeq
This implies that the lower bounds on the fractional energies depend
solely on the individual parton identities, rather than on the splitting 
types.  We emphasise, however, that in principle the cutoffs in 
eqs.~(\ref{thetapm}) are not limited to a particular perturbative order.

In practice, the cutoffs are flavour- and scale-dependent parameters
of a particular MCEG. In the angular-ordered parton shower formalism of
ref.~\cite{Gieseke:2003rz}, the shower evolution scale $\mu$ is the
variable called $\tilde q$ there, and the cutoffs are functions of this 
variable and the effective parton masses $m_i$, which at high scales 
approximate to eq.~(\ref{epflavLO}) with $\epsilon_i= m_i/\tilde q$. In a 
dipole shower formalism such as ref.~\cite{Schumann:2007mg}, the shower 
evolution variable is the transverse momentum of emission relative to the 
dipole and the cutoffs are functions of this variable, the effective parton 
masses and the minimum resolvable transverse momentum. It should be
emphasised that these cutoffs, which apply throughout backward
showering and concern the resolvability of emitted timelike partons,
are distinct from the cutoff scale at which backward evolution
is terminated and the evolving spacelike parton is merged into an
incoming hadron.

In keeping with what has been done so far, the quantities defined
in eq.~(\ref{thetapm}) can be arranged compactly in two matrices,
$\Top_z^\Rin$ and $\Top_z^\Rout$, whose elements are:
\beq
\left(\Top_z^\Rin\right)_{ij}=\stepf_{ij,z}^\Rin\,,\;\;\;\;\;\;\;\;
\left(\Top_z^\Rout\right)_{ij}=\stepf_{ij,z}^\Rout\,.
\label{Topdef}
\eeq
For any function $g(z)$ and pair of parton indices $(i,j)$, we can exploit 
the following identity:
\beqn
\left[g(z)\right]_+ &=& 
\left[g(z)\,\stepf_{ij,z}^\Rout\right]_+ + 
\left[g(z)\,\stepf_{ij,z}^\Rin\right]_+ 
\nonumber\\*&=&
\left[g(z)\,\stepf_{ij,z}^\Rout\right]_+ + 
\left(g(z)\,\stepf_{ij,z}^\Rin\right) +
\left(-\int_0^1 d\omega\,g(\omega)\stepf_{ij,\omega}^\Rin\right)\delta(1-z)\,,
\phantom{aaa}
\label{fplusid}
\eeqn
and rewrite eq.~(\ref{Oopform}) as follows:
\beq
\Oop(z)=\left[\Aop(z)\circ\Top_z^\Rout\right]_+ + \Aop(z)\circ\Top_z^\Rin
+\overline{\Bop}\,\delta(1-z) + 
\Cop(z)\circ\Top_z^\Rout+ \Cop(z)\circ\Top_z^\Rin\,,
\label{Oopform3}
\eeq
where by $\circ$ we have denoted the element-by-element matrix 
multiplication, e.g.:
\beq
\left(\Aop\circ\Top\right)_{ij}=
\left(\Aop\right)_{ij}\left(\Top\right)_{ij}
\eeq
and:
\beq
\overline{\Bop}=\Bop-\int_0^1 dz\,\Aop(z)\circ\Top_z^\Rin\,.
\label{bBop}
\eeq
For those operators for which eq.~(\ref{ABopid}) holds,
eq.~(\ref{bBop}) can be written in the equivalent form\footnote{We
point out that eq.~(\ref{Oopform3}) is unchanged. This implies, 
in particular, that $\widetilde{\Aop}(z)/(1-z)$ is inside the plus
prescription in the first term on the r.h.s..\label{ft:subtr}}:
\beq
\overline{\Bop}=\widetilde{\Bop}+
\int_0^1 dz\,\frac{\widetilde{\Aop}(z)-\widetilde{\Aop}(1)}{1-z}
\circ\Top_z^\Rout
-\int_0^1 dz\,\frac{\widetilde{\Aop}(1)\circ\Top_z^\Rin}{1-z}\,.
\label{bBop2}
\eeq
With eq.~(\ref{Oopform3}), we can write the evolution equations as follows:
\beq
\frac{\partial F(x)}{\partial\log\mut}=
\Oop\otimes_x F=\Wop\left[F\right](x)+\Zop\left[F\right](x)+
\overline{\Bop}F(x)\,,
\label{evolmu3}
\eeq
where:
\beqn
\Wop\left[F\right](x)&=&\Big(\big[\Aop\circ\Top^\Rout\big]_+
+\Cop\circ\Top^\Rout\Big)\otimes_x F\,,
\label{WJop}
\\
\Zop\left[F\right](x)&=&\Big(\big[\Aop+\Cop\big]\circ\Top^\Rin\Big)
\otimes_x F\,.
\label{ZKop}
\eeqn
With the scalar functions introduced in eq.~(\ref{AopBopCop}),
these are ($(F)_i=f_i$):
\beqn
\Big(\big[\Aop\circ\Top^\Rout\big]_+\otimes_x F\Big)_i&\!\!=\!\!&
\int_0^1 dz\,A_i(z)\stepf_{ii,z}^\Rout\left[\frac{\stepf(z\ge x)}{z}
f_i\left(\frac{x}{z}\right)-f_i(x)\right],\phantom{aa}
\\
\Big(\big(\Aop\circ\Top^\Rin\big)\otimes_x F\Big)_i&\!\!=\!\!&
\int_0^1 dz\,A_i(z)\stepf_{ii,z}^\Rin\,\frac{\stepf(z\ge x)}{z}
f_i\left(\frac{x}{z}\right),
\\
\Big(\big(\Cop\circ\Top^{\Rinout}\big)\otimes_x F\Big)_i&\!\!=\!\!&\sum_j
\int_0^1 dz\,C_{ij}(z)\stepf_{ij,z}^{\Rinout}\,\frac{\stepf(z\ge x)}{z}
f_j\left(\frac{x}{z}\right).
\eeqn
By construction, the r.h.s.~of eq.~(\ref{evolmu3}) is cutoff-independent.
One can show that the contribution to $(\Wop\left[F\right](x))_i$ from the 
splitting $j\to ik$ is power-suppressed when \mbox{$x<1-\ep_{j\ominus i}$}.
Overall, $(\Wop\left[F\right](x))_i$ cannot be power-suppressed when 
\mbox{$x>1-\ep$}, with \mbox{$\ep=\min_j\ep_{j\ominus i}$}, because in that 
region $(\Zop\left[F\right](x))_i=0$, since for such $x$ values one has 
$\stepf_{ij,z}^\Rin\stepf(z\ge x)=0$ for any $z$ and $j$. Therefore,
the cutoff dependence of $\Wop\left[F\right](x)$ must cancel that
of the $\overline{\Bop}F(x)$ term, which is in general logarithmic (see 
e.g.~eq.~(\ref{bBop})). From a physical viewpoint, $\Zop\left[F\right]$
describes resolved (owing to $\Top^\Rin$) real emissions with 
\mbox{$\max(x,\ep)\le z\le 1-\ep$}, while $\overline{\Bop}F(x)$ describes 
virtual emissions (being proportional to $F(x)$). The term $\Wop\left[F\right]$ 
is a remainder\footnote{This distinction between $\Zop$ and $\Wop$ is
not entirely precise, owing to the possible flavour dependence of the
cutoffs, which implies that certain kinematical configurations are
resolvable only for certain types of branchings. The underpinning physical 
picture is nevertheless correct.}
that arises from the fact that the kernels of the evolution 
equations are not ordinary functions, but distributions that involve 
subtractions; from a physics viewpoint, it may be associated with
branchings resolvable in the scale but not in Bjorken $x$.

With MC applications in mind, eq.~(\ref{evolmu3}) can be further
manipulated by writing:
\beq
\overline{\Bop}(\mut)=\frac{\mut}{\Sop(\mut)}\,
\frac{\partial\Sop(\mut)}{\partial\mut}\,,
\label{bBvssud}
\eeq
with
\beq
\left(\Sop(\mut)\right)_{ij}=\delta_{ij}S_i(\mut)\,,\;\;\;\;\;\;\;\;
\frac{1}{\Sop}\equiv\left(\Sop\right)^{-1}\;\;\;\Longrightarrow\;\;\;
\left(\frac{1}{\Sop}\right)_{ij}=\delta_{ij}\frac{1}{S_i}\,,
\eeq
and
\beq
S_i(\mut)=\exp\left[\int_{\muzt}^{\mut}\frac{d\kappa^2}{\kappa^2}
\overline{B}_i(\kappa^2)\right]
\;\;\;\;\Longleftrightarrow\;\;\;\;
\Sop(\mut)=\exp\left[\int_{\muzt}^{\mut}\frac{d\kappa^2}{\kappa^2}
\overline{\Bop}(\kappa^2)\right].
\label{Sidef}
\eeq
In other words, $\Sop$ is a diagonal matrix that collects the Sudakov
form factors. As such, it may seem that the sign in the exponent in 
eq.~(\ref{Sidef}) is the opposite w.r.t.~the standard one, but in fact
this is not the case, as can be understood from eq.~(\ref{bBop}).
With eq.~(\ref{bBvssud}), eq.~(\ref{evolmu3}) can be cast as follows:
\beq
\frac{\partial}{\partial\mut}\left(\frac{1}{\Sop(\mut)}F(\mut)\right)=
\frac{1}{\mut\,\Sop(\mut)}\,
\Big(\Wop\left[F\right](\mut)+\Zop\left[F\right](\mut)\Big)\,,
\eeq
which can be put in an integrated form, thus:
\beq
F(\mut)=\frac{\Sop(\mut)}{\Sop(\muzt)}\,F(\muzt)+
\int_{\muzt}^{\mut}\frac{d\kappa^2}{\kappa^2}
\frac{\Sop(\mut)}{\Sop(\kappa^2)}\,
\Big(\Wop\left[F\right](\kappa^2)+\Zop\left[F\right](\kappa^2)\Big)\,,
\label{intevolmu}
\eeq
or, alternatively, thus\footnote{The r.h.s.~of eq.~(\ref{intevolexpmu})
features the multiplication of two column vectors, which is meant as
an element-by-element multiplication. Since no confusion is possible with
the multiplications that feature the transpose of a column vector, no
special symbol has been introduced here.}:
\beq
\frac{\Sop(\mut)}{\Sop(\muzt)}\,\frac{F(\muzt)}{F(\mut)}=
\exp\left[-\int_{\muzt}^{\mut}\frac{d\kappa^2}{\kappa^2}
\frac{1}{F(\kappa^2)}\,
\Big(\Wop\left[F\right](\kappa^2)+\Zop\left[F\right](\kappa^2)\Big)\right]\,.
\label{intevolexpmu}
\eeq
We stress again that eqs.~(\ref{intevolmu}) and~(\ref{intevolexpmu})
are fully equivalent to eq.~(\ref{evolmu3}) but, being in an integrated
form, they also include the information on the initial conditions ($F(\muzt)$).
In turn, they are all equivalent to the original evolution equation,
eq.~(\ref{evolmu2}). Thus, in spite of the fact that they feature
cutoff-dependent kernels ($\Bop$, $\Zop$, and $\Wop$), the PDFs that
solve them are cutoff-independent. In fact, if one were interested
only in determining the PDFs, the solution of eq.~(\ref{evolmu2})
(best obtained in Mellin space) would be much more straightforward
than that of eqs.~(\ref{intevolmu}) or~(\ref{intevolexpmu}). The
primary interest of the latter equations is in the fact that they
are expressed in terms of the same quantities that are used in
initial-state parton showers; as such, they can be regarded as
giving consistency conditions among these quantities that initial-state 
parton showers (which assume knowledge of the PDFs) must respect.
We shall show later that, in the context of the current approaches used
in MCs, this is not quite the case.

\subsection{Ambiguities and choices\label{sec:ref}} 
One of the ingredients of the manipulation of the PDF evolution
equations is the definition of the Sudakov form factors. We point
out that, by means of eq.~(\ref{bBvssud}), we have defined them by
exponentiating the entire virtual term that appears in eq.~(\ref{evolmu3}). 
This is not mandatory, and in fact it may lead to problems (see
e.g.~sect.~\ref{sec:PDFcut}). In the context of a more flexible
approach, we start by writing the rightmost term 
on the r.h.s.~of eq.~(\ref{evolmu3}) as follows:
\beq
\overline{\Bop}^\Rout F(x)+\overline{\Bop}^\Rin F(x)\,,
\label{Bopsplit}
\eeq
for any two quantities $\overline{\Bop}^\Rout$ and $\overline{\Bop}^\Rin$ 
such that:
\beq
\overline{\Bop}=\overline{\Bop}^\Rout+\overline{\Bop}^\Rin\,.
\label{Bopsplit2}
\eeq
Then, we define the Sudakov factors thus
\beq
\Sop(\mut)=\exp\left[\int_{\muzt}^{\mut}\frac{d\kappa^2}{\kappa^2}
\overline{\Bop}^\Rin(\kappa^2)\right],
\label{Sidef2}
\eeq
rather than with eq.~(\ref{Sidef}), and we include the contribution
from $\overline{\Bop}^\Rout$ in the $\Wop[F]$ functional.
In order to do that, and also in view of future use (see also
appendix~\ref{sec:PDFsol}), it turns out to be convenient to introduce 
the two evolution operators:
\beqn
\Oop^\Rout(z)&=&\left[\Aop(z)\circ\Top_z^\Rout\right]_+ + 
\overline{\Bop}^\Rout\,\delta(1-z) + \Cop(z)\circ\Top_z^\Rout\,,
\label{Oopgt}
\\
\Oop^\Rin(z)&=&\Aop(z)\circ\Top_z^\Rin + \overline{\Bop}^\Rin\,\delta(1-z) + 
\Cop(z)\circ\Top_z^\Rin\,,
\label{Ooplt}
\eeqn
which, loosely speaking, account for emissions in the outer
(unresolved) and inner (resolved) regions, respectively. By
construction (see eq.~(\ref{Oopform3})):
\beq
\Oop(z)=\Oop^\Rout(z)+\Oop^\Rin(z)
\;\;\;\;\Longrightarrow\;\;\;\;
\frac{\partial F(x)}{\partial\log\mut}=
\Oop^\Rout\otimes_x F+\Oop^\Rin\otimes_x F\,,
\label{evolmu4}
\eeq
and
\beqn
\Wop\left[F\right](x)&=&\Oop^\Rout\otimes_x F
\label{OopgtW}
\\*&=&
\Big(\big[\Aop\circ\Top^\Rout\big]_+
+\Cop\circ\Top^\Rout\Big)\otimes_x F+\overline{\Bop}^\Rout F(x)\,,
\label{WJop2}
\\
\Zop\left[F\right](x)+\overline{\Bop}^\Rin F(x)&=&\Oop^\Rin\otimes_x F\,.
\label{OopltZB}
\eeqn
As was anticipated, owing to eq.~(\ref{Bopsplit2}) the expression
of $\Wop[F]$ in eq.~(\ref{WJop2}) is in general not the same as that
in eq.~(\ref{WJop}), while that of $\Zop[F]$ is still given by 
eq.~(\ref{ZKop}). The crucial thing is that, by taking into account the
redefinition of the Sudakov factor and of the $\Wop[F]$ functional, the 
integrated form of the evolution equation is still given by 
eq.~(\ref{intevolmu}) or eq.~(\ref{intevolexpmu}).

While eq.~(\ref{Bopsplit}) is so far largely arbitrary, given the 
interpretation of $\Wop$ it is wise to require that:
\beq
\lim_{\ep\to 0}\overline{\Bop}^\Rout=0\,,
\;\;\;\;\;\;\;\;
\ep=\{\epLij,\epUij\}_{ij}\,.
\label{limBgt}
\eeq
This constraint implies that the Sudakov form factors that one would obtain 
by choosing different $\overline{\Bop}^{\Rinout}$ would differ from one 
another by terms suppressed by powers of the cutoffs. In terms of the 
quantities that appear in eq.~(\ref{Oopform}), the above can be rewritten 
by exploiting eq.~(\ref{bBop}), thus:
\beq
\overline{\Bop}^\Rout=\Bop^\Rout\,,\;\;\;\;\;\;
\overline{\Bop}^\Rin=\Bop^\Rin-\int_0^1 dz\,\Aop(z)\circ\Top_z^\Rin\,,
\label{bBoplg}
\eeq
with
\beq
\Bop=\Bop^\Rout+\Bop^\Rin\,,\;\;\;\;\;\;\;\;
\lim_{\ep\to 0}\Bop^\Rout=0\,.
\label{Bopaux}
\eeq
We stress that associating the {\em entire} second term on the r.h.s.~of
eq.~(\ref{bBop}) with $\overline{\Bop}^\Rin$ is merely a sensible choice,
but a choice nevertheless. For example, we could have associated a
\mbox{$(1-\ep)$} fraction of it with $\overline{\Bop}^\Rin$, and the
remaining $\ep$ fraction with $\overline{\Bop}^\Rout$. In the following,
we shall not exploit this option, and always employ eqs.~(\ref{bBoplg}) 
and~(\ref{Bopaux}), so that the flexibility in choosing 
$\overline{\Bop}^{\Rinout}$ will be entirely controlled by the choice
of $\Bop^{\Rinout}$.

In the cases where eq.~(\ref{ABopid}) holds,
eqs.~(\ref{Oopgt}) and~(\ref{Ooplt}) become:
\beqn
\Oop^\Rout(z)&=&
\left[\frac{\widetilde{\Aop}(z)}{1-z}\circ\Top_z^\Rout\right]_+ + 
\overline{\Bop}^\Rout\,\delta(1-z) + \Cop(z)\circ\Top_z^\Rout\,,
\label{Oopgt2}
\\
\Oop^\Rin(z)&=&\frac{\widetilde{\Aop}(z)}{1-z}\circ\Top_z^\Rin + 
\overline{\Bop}^\Rin\,\delta(1-z) + \Cop(z)\circ\Top_z^\Rin\,,
\label{Ooplt2}
\eeqn
where, by taking eq.~(\ref{bBop2}) into account:
\beq
\overline{\Bop}^\Rout=\widetilde{\Bop}^\Rout+
\int_0^1 dz\,\frac{\widetilde{\Aop}(z)-\widetilde{\Aop}(1)}{1-z}
\circ\Top_z^\Rout
\,,\;\;\;\;\;\;
\overline{\Bop}^\Rin=\widetilde{\Bop}^\Rin
-\int_0^1 dz\,\frac{\widetilde{\Aop}(1)\circ\Top_z^\Rin}{1-z}\,,
\label{bBop2lg}
\eeq
with
\beq
\widetilde{\Bop}=\widetilde{\Bop}^\Rout+\widetilde{\Bop}^\Rin\,,
\;\;\;\;\;\;\;\;
\lim_{\ep\to 0}\widetilde{\Bop}^\Rout=0\,.
\label{wBopaux}
\eeq
Here, the same remark made after eq.~(\ref{Bopaux}) applies: namely,
the association of the two rightmost terms of eq.~(\ref{bBop2}) 
with $\overline{\Bop}^\Rout$ and $\overline{\Bop}^\Rin$, respectively, 
as is done in eq.~(\ref{bBop2lg}) is a choice we shall always adhere to, 
and for the operators of this form the flexibility in choosing 
$\overline{\Bop}^{\Rinout}$ will be controlled by the choice
of $\widetilde{\Bop}^{\Rinout}$.

There is an easy way to enforce the conditions in eqs.~(\ref{Bopaux}) 
and~(\ref{wBopaux}) that is, once again, quite arbitrary, but that
allows an easy interpretation from a physical viewpoint, and leads to 
the Sudakov form factors which are typically adopted at the LO in QCD 
(see sect.~\ref{sec:LO}). Namely, one finds functions $b_{ij}(\omega)$ 
and $\tilde{b}_{ij}(\omega)$ which are bounded from above
and below, and are such that:
\beq
B_j=\int_0^1 d\omega\sum_i b_{ij}(\omega)\,,
\;\;\;\;\;\;\;\;
\widetilde{B}_j=\int_0^1 d\omega\sum_i\tilde{b}_{ij}(\omega)\,,
\label{BtBlocal}
\eeq
and defines:
\beq
B^{\Rinout}_j=\int_0^1 d\omega\sum_i b_{ij}(\omega)\,
\stepf_{ij,\omega}^{\Rinout}\,,
\;\;\;\;\;\;\;\;
\widetilde{B}^{\Rinout}_j=\int_0^1 d\omega\sum_i\tilde{b}_{ij}(\omega)\,
\stepf_{ij,\omega}^{\Rinout}\,.
\label{BtBgllocal}
\eeq
Note that in the case where eq.~(\ref{ABopid}) holds, $\Bop$ (and
therefore the functions $b_{ij}(\omega)$) need not necessarily be introduced.
If one still finds it convenient to do so (e.g.~to use both the form
of eq.~(\ref{Oopform}) and that of eq.~(\ref{Oopform2})), eqs.~(\ref{ABopid}) 
and~(\ref{BtBlocal}) imply:
\beq
\int_0^1 d\omega\,b_{ii}(\omega)=
\int_0^1 d\omega\,\tilde{b}_{ii}(\omega)+
\int_0^1 dz\,\frac{\widetilde{A}_i(z)-\widetilde{A}_i(1)}{1-z}\,.
\label{bvstbint}
\eeq
Clearly, the easiest way to achieve this and be consistent with
eqs.~(\ref{bBoplg})--(\ref{wBopaux}) is to work with a local version
of eq.~(\ref{bvstbint}), namely:
\beq
b_{ii}(z)=\tilde{b}_{ii}(z)+
\frac{\widetilde{A}_i(z)-\widetilde{A}_i(1)}{1-z}\,.
\label{bvstblocal}
\eeq
We anticipate that at the LO in QCD the choice of the $b_{ij}(\omega)$ and 
$\tilde{b}_{ij}(\omega)$ functions along the lines presented above leads to
Sudakov form factors expressed as integrals of the LO splitting kernels
(see sect.~\ref{sec:LO} for more details). However, this discussion should
render it clear that, even at the LO in QCD, this is a choice that is
not dictated by any fundamental principle, but by convenience and
ease of interpretation.

The separation of the virtual terms in eq.~(\ref{Bopsplit}) stemming 
from eq.~(\ref{Bopsplit2}) encompasses the case where such a separation 
is not considered. Even after making a definite choice for the cutoffs,
one can continuously pass from one scenario to the other by means of 
the replacements\footnote{Although in general the parameter $\lambda$ can 
be flavour-dependent, for our purposes such a dependence can be neglected.}
\beq
\overline{\Bop}^\Rout\;\longrightarrow\;\lambda\overline{\Bop}^\Rout\,,
\;\;\;\;\;\;\;\;
\overline{\Bop}^\Rin\;\longrightarrow\;
\overline{\Bop}-\lambda\overline{\Bop}^\Rout\,,
\label{lamrepl}
\eeq
with $0\le\lambda\le 1$ in all quantities that feature a dependence
on $\overline{\Bop}^\Rout$ and/or $\overline{\Bop}^\Rin$.

\section{Monte Carlo backward evolution\label{sec:MCbw}}
When an MC generates initial-state parton showers, the PDFs are thought 
to be given: they are employed to ``guide'' the backward evolution.
One usually assumes that consistency demands that the 
longitudinal momentum left after all branchings have occurred be distributed 
according to the given PDFs (this identification holds in a statistical sense; 
it is exact only after an infinite number of showers have been carried out). 
However, since only {\em resolved} branchings (i.e., those with 
\mbox{$\ep<z<1-\ep$}) can be generated, the identification above can be true 
only in the resolved region. In fact, as we shall show, even in the resolved
region MCs are generally not able to reconstruct the PDFs. Ultimately,
this arises from the fact that the PDF evolution equations are expressed
as convolution integrals, and thus the derivative w.r.t.~the scale of the 
PDF at a given $x$ receives contributions from all $z$'s, with $x\le z\le 1$.
In other words, the unresolved region feeds into the resolved region
as well as itself. This is
unavoidable: for PDF evolution, the separation between the resolved
and unresolved regions is totally arbitrary, and has no bearing on
the final form of the PDFs.

Conversely, MCs cannot function without a clear separation between
resolved and unresolved regions, i.e.~without the introduction of
cutoffs. As is well known, this leads to the possibility of generating
showers by means of an iterative Markovian process, one of whose key
ingredients is the inversion of the so-called non-emission probability 
(NEP henceforth), which gives one the scale at which the next parton
branching occurs. The usual argument adopted for deriving the NEP
associated with initial-state emissions exploits a partonic picture 
of the PDFs, whereby these ``count'' the number of partons at any given 
values of the Bjorken $x$ and scale. As was said before, one identifies 
the $\Zop$ and $\Wop$ contributions to the PDF evolution as associated 
with resolvable branchings and with branchings resolvable in $\mu$ but 
not in $z$, respectively. Therefore, the number of partons of flavour $i$ 
that do not undergo branchings of any type in the range \mbox{$(\muzt,\mut)$} 
is equal to:
\beq
\frac{S_i(\mut)}{S_i(\muzt)}\,f_i(x,\muzt)\,,
\label{numNBwoJ}
\eeq
while that of partons that either do not branch, or branch in a manner
unresolvable in $x$, is equal to:
\beq
\frac{S_i(\mut)}{S_i(\muzt)}\,f_i(x,\muzt)+
\int_{\muzt}^{\mut}\frac{d\kappa^2}{\kappa^2}
\frac{S_i(\mut)}{S_i(\kappa^2)}\,
\big(\Wop\left[F\right]\!\big)_i(x,\kappa^2)\,.
\label{numNBJ}
\eeq
The NEP is defined as the fraction of partons that do not branch in a
resolvable manner between any two scales. The difference between
forward and backward evolution is simply the reference relative
to which that fraction is measured, because the elementary 
branching mechanism must not be affected by the direction of the evolution.
For an evolution in the range \mbox{$(\muzt,\mut)$}, if the evolution is
forwards (backwards) the reference is $f_i(x,\muzt)$ ($f_i(x,\mut)$).
Thus, eq.~(\ref{numNBJ}) leads to the non-emission probabilities
for the forward and backward evolution of a parton of type $i$,
\beqn
&\!\!\!\!\mbox{Forward:} &\phantom{aaa}{\rm NEP}_i=
\frac{S_i(\mut)}{S_i(\muzt)}+
\frac{1}{f_i(x,\muzt)}\,\int_{\muzt}^{\mut}\frac{d\kappa^2}{\kappa^2}
\frac{S_i(\mut)}{S_i(\kappa^2)}\,
\big(\Wop\left[F\right]\!\big)_i(x,\kappa^2),
\label{FnepJ}
\\
&\!\!\!\!\mbox{Backward:} &\phantom{aaa}{\rm NEP}_i=
\frac{S_i(\mut)}{S_i(\muzt)}\,\frac{f_i(x,\muzt)}{f_i(x,\mut)}+
\frac{1}{f_i(x,\mut)}\,\int_{\muzt}^{\mut}\frac{d\kappa^2}{\kappa^2}
\frac{S_i(\mut)}{S_i(\kappa^2)}\,
\big(\Wop\left[F\right]\!\big)_i(x,\kappa^2).\phantom{aaaa}
\label{BnepJ}
\eeqn
Here we are concerned with the backward case, which will
henceforth always be implied.  Then from eqs.~(\ref{BnepJ}) and
(\ref{intevolmu}) one also obtains:
\beq
\NEPbwi=
1-\frac{1}{f_i(x,\mut)}\,\int_{\muzt}^{\mut}\frac{d\kappa^2}{\kappa^2}
\frac{S_i(\mut)}{S_i(\kappa^2)}\,
\big(\Zop\left[F\right]\!\big)_i(x,\kappa^2)\,,
\label{BnepJwZ}
\eeq
consistently with the meaning of the $\Zop[F]$ functional. In a backward
evolution, which proceeds from larger to smaller scales, starting from
a given $\mut$ one obtains the ``next'' scale $\muzt<\mut$ by solving 
for $\muzt$ the equation
\beq
r=\NEPbwi\,,
\label{BEstep}
\eeq
with $0<r<1$ a uniform random number, and $i$ given. After selecting
the branching channel and its momentum fraction, the procedure is 
iterated until a $\muzt$ value is obtained that is smaller than some 
pre-defined threshold (the so-called hadronization scale). However, 
MCEGs do not literally solve eq.~(\ref{BEstep}), but
either~\cite{Marchesini:1987cf}
\beq
r=\NEPbwiR\equiv
\frac{S_i(\mut)}{S_i(\muzt)}\,\frac{f_i(x,\muzt)}{f_i(x,\mut)}\,,
\label{BEstepR}
\eeq
or~\cite{Sjostrand:1985xi}
\beq
r=\NEPbwiE\equiv
\exp\left[-\int_{\muzt}^{\mut}\frac{d\kappa^2}{\kappa^2}
\frac{1}{f_i(x,\kappa^2)}\,
\big(\Zop\left[F\right]\big)_i(x,\kappa^2)\right]\,,
\label{BEstepE}
\eeq
where the superscript R or E indicates that a ratio or an exponential
approximation for the NEP has been used, respectively.
The crucial thing, which follows directly from the evolution equations 
as given in eq.~(\ref{intevolmu}) or eq.~(\ref{intevolexpmu}), is that:
\beq
\mbox{if}\phantom{aaaa}\Wop\left[F\right]=0
\phantom{aaaa}\mbox{then}\phantom{aaaa}
\NEPbwi=
\NEPbwiR=
\NEPbwiE\,.
\label{NEPWeq0}
\eeq
Thus, in the resolved region the solution of eq.~(\ref{BEstep}) coincides 
with that of eqs.~(\ref{BEstepR}) or~(\ref{BEstepE}) up to terms suppressed 
by powers of the cutoffs (since in that region $\Wop$ vanishes with
the cutoffs). This is the reason why, in standard current practice,
eq.~(\ref{BEstepR}) and eq.~(\ref{BEstepE}) are considered equivalent
to one another -- effects suppressed by powers of the cutoffs are
systematically neglected. This is in fact a dangerous position to take,
given that MC cutoffs are often not particularly small, their effects can 
accumulate over the course of evolution, and there is no other mechanism 
that forces $\Wop$ to vanish bar the vanishing of the cutoffs.

It is therefore instructive to see how the three NEP expressions
considered above differ from each other when power-suppressed cutoff
effects are not neglected. We start by considering the probability
distribution of the scale of the next branching which, according
to eqs.~(\ref{BEstep}), (\ref{BEstepR}), and~(\ref{BEstepE}), is
given by the derivative w.r.t.~$\muzt$ of the respective NEP. By direct 
computation, and by employing the evolution equations~(\ref{intevolmu})
and~(\ref{intevolexpmu}), we obtain:
\beqn
&&\frac{\partial}{\partial\log\muzt}\,\NEPbwi=
\frac{1}{f_{i}(x,\mut)}\frac{S_{i}(\mut)}{S_{i}(\muzt)}\,
\big(\Zop\left[F\right]\!\big)_{i}(x,\muzt)\,,
\label{dNEPdl0}
\\
&&\frac{\partial}{\partial\log\muzt}\,\NEPbwiR=
\frac{1}{f_{i}(x,\mut)}\frac{S_{i}(\mut)}{S_{i}(\muzt)}
\Big[\big(\Wop\left[F\right]\!\big)_{i}(x,\muzt)+
\big(\Zop\left[F\right]\!\big)_{i}(x,\muzt)\Big],
\label{dNEPrdk0}
\\
&&\frac{\partial}{\partial\log\muzt}\,\NEPbwiE=
\label{dNEPEdl0}
\\*&&\phantom{aaaaaa}
\frac{1}{f_{i}(x,\mut)}\frac{S_{i}(\mut)}{S_{i}(\muzt)}\,
\big(\Zop\left[F\right]\!\big)_{i}(x,\muzt)\,
\exp\!\left[\int_{\muzt}^{\mut}\frac{d\kappa^2}{\kappa^2}
\frac{1}{f_{i}(x,\kappa^2)}\,
\big(\Wop\left[F\right]\!\big)_{i}(x,\kappa^2)\right].
\nonumber
\eeqn
Equation~(\ref{dNEPdl0}), since it factors out $\Zop$ that is non-null
only for resolved emissions, shows that $\NEPbwi$ is consistent with the
requirement that the NEP be associated with the fraction of partons that
do not branch in a resolvable manner. This may seem to be the case also
for $\NEPbwiE$, but in fact the exponentiated $\Wop$ term in 
eq.~(\ref{dNEPEdl0}) introduces a spurious extra cutoff dependence w.r.t.~the 
evolution generated by means of $\NEPbwi$. Finally, in eq.~(\ref{dNEPrdk0})
the $\Wop$ and $\Zop$ contributions are on the same footing: this is
because, as the comparison between eqs.~(\ref{numNBwoJ}) and~(\ref{numNBJ})
shows, $\NEPbwiR$ is actually the NEP for no branchings, regardless whether
they are resolved or unresolved in $x$.

In appendix~\ref{sec:PDFsol} we discuss in detail the implications 
of eqs.~(\ref{dNEPdl0})--(\ref{dNEPEdl0}) for the requirement that
MC backward evolution allows one to reconstruct the PDFs given in
input to the parton shower. The bottom line is that, in practice, 
such a reconstruction always fails. It can be made to {\em formally} 
succeed with $\NEPbwiR$, while if $\NEPbwi$ is adopted one can 
reconstruct PDFs where all non-resolved contributions are consistently
neglected; the same is true for $\NEPbwiE$ if a branching-by-branching
reweighting is applied.

The above suggests that $\NEPbwi$ and $\NEPbwiE$ are closer to each
other than either is to $\NEPbwiR$. This can be also seen in another way, 
by considering the differences between any two of these quantities.
From eqs.~(\ref{BnepJ}) and~(\ref{BEstepR}) we obtain:
\beq
\NEPbwiR-\NEPbwi=
-\frac{S_i(\mut)}{f_i(\mut)}\,
\int_{\muzt}^{\mut}\frac{d\kappa^2}{\kappa^2}
\frac{1}{S_i(\kappa^2)}\,
\big(\Wop\left[F\right]\!\big)_i(\kappa^2)\equiv\ord(\as)\,,
\label{diffNEPR}
\eeq
whereas from eqs.~(\ref{BnepJ}) and~(\ref{BEstepE}):
\beqn
&&\NEPbwiE-\NEPbwi
\nonumber\\*&&\phantom{aa}=
\frac{S_i(\mut)}{f_i(\mut)}\,
\int_{\muzt}^{\mut}\frac{d\kappa^2}{\kappa^2 S_i(\kappa^2)}
\left(\frac{f_i(\muzt)}{f_i(\kappa^2)}\,\frac{S_i(\kappa^2)}{S_i(\muzt)}
-1\right)\big(\Wop\left[F\right]\!\big)_i(\kappa^2)
\nonumber\\*&&\phantom{aa=}+
\half\,\frac{f_i(\muzt)}{f_i(\mut)}\,\frac{S_i(\mut)}{S_i(\muzt)}
\left(\int_{\muzt}^{\mut}\frac{d\kappa^2}{\kappa^2}
\frac{1}{f_i(\kappa^2)}\,
\big(\Wop\left[F\right]\!\big)_i(\kappa^2)\right)^2+\ldots
\nonumber
%%\label{diffNEPE1}
\\*&&\phantom{aa}=
-\frac{S_i(\mut)}{f_i(\mut)}\,
\int_{\muzt}^{\mut}\frac{d\kappa^2}{\kappa^2 f_i(\kappa^2)}
\int_{\muzt}^{\kappa^2}\frac{d\rho^2}{\rho^2 S_i(\rho^2)}
\Big(\big(\Wop\left[F\right]\!\big)_i(\rho^2)+
\big(\Zop\left[F\right]\!\big)_i(\rho^2)\Big)
\big(\Wop\left[F\right]\!\big)_i(\kappa^2)
\nonumber\\*&&\phantom{aa=}+
\half\,\frac{f_i(\muzt)}{f_i(\mut)}\,\frac{S_i(\mut)}{S_i(\muzt)}
\left(\int_{\muzt}^{\mut}\frac{d\kappa^2}{\kappa^2}
\frac{1}{f_i(\kappa^2)}\,
\big(\Wop\left[F\right]\!\big)_i(\kappa^2)\right)^2+\ldots
\label{diffNEPE2}
\\*&&\phantom{aa}\equiv
\ord(\as^2)\,,
\label{diffNEPE3}
\eeqn
where the ellipsis represents terms with three or more $\Wop$ terms,
and we have used eqs.~(\ref{intevolmu}) and~(\ref{intevolexpmu}).
The powers of $\as$ in eqs.~(\ref{diffNEPR}) and~(\ref{diffNEPE3})
stem from having regarded both the PDFs and the Sudakovs as 
quantities of perturbative $\ord(1)$, while both $\Wop$ and $\Zop$ are
of $\ord(\as)$ (see eqs.~(\ref{WJop}) and~(\ref{ZKop})).

We finally note that, when power-suppressed effects are not
neglected, the simple probabilistic interpretation upon which
MCs rely to perform initial-state backward evolution may lose
validity. In all cases, this can be seen to come from the fact
that $\Wop[F]$ has no definite sign. Thus, from eq.~(\ref{intevolexpmu})
one sees that $\NEPbwiR$ is not necessarily monotonic, and from
eq.~(\ref{intevolmu}) that $\NEPbwi$ is not necessarily positive.
In both cases, this implies that, in some regions of the phase space
(typically, at large Bjorken $x$), these NEPs are actually not cumulative
probability distributions, and therefore that the solution of 
eq.~(\ref{BEstepR}) or eq.~(\ref{BEstep}) may not exist, or may not 
be unique. As far as $\NEPbwiE$ is concerned, it is positive definite, 
monotonic, and bounded by one; however, as was discussed in relation 
to eq.~(\ref{dNEPEdl0}), its physical interpretation is unclear.

We conclude by remarking that, while $\NEPbwi$ may turn out to be 
negative, it generally is positive. One can in fact turn the requirement
that it be positive into a tool to determine the cutoffs in a 
physically-meaningful manner, in the sense of limiting the
impact of non-resolvable emissions to an extent that allows one to
recover a probabilistic interpretation.

Another way to approach the problem is to acknowledge the fact
that PDFs and initial-state parton showers are inherently incompatible
at some level, and to construct MC-specific PDFs by means of which
all issues are removed {\em ab initio}. This option will be discussed
in sect.~\ref{sec:PDFcut}.

\section{The LO QCD case\label{sec:LO}}
The general approach of sects.~\ref{sec:gen} and~\ref{sec:MCbw} can be 
applied to the case which is currently the most relevant to MC simulations, 
namely that where only the LO evolution kernels are considered.
In order to simplify our discussion, we ignore complications due to
different quark masses; thus, we shall not need to specify the individual
flavours, but only generic cutoffs $\epsilon_q$ and $\epsilon_g$ for
quark and gluon emission, respectively.

We denote the LO kernels, i.e.~the elements of $\APmat^{[0]}$
in eq.~(\ref{Oopform2}), as follows\footnote{Bearing in mind that
at the LO there are no $q\bq$ kernels, in the notation we need not
distinguish quarks and antiquarks.\label{ft:qvsbq}}:
\beqn
P_{qq}(z)&=&\CF\left(\frac{1+z^2}{1-z}\right)_+\,,
\label{Pqq}
\\
P_{gq}(z)&=&\CF\,\frac{1+(1-z)^2}{z}\,,
\label{Pgq}
\\
P_{qg}(z)&=&\TF\left(z^2+(1-z)^2\right)\,,
\label{Pqg}
\\
P_{gg}(z)&=&2\CA\left(\frac{z}{(1-z)_+}+\frac{1-z}{z}+z(1-z)\right)
+\gamma(g)\delta(1-z)\,,
\label{Pgg}
\eeqn
with ($\NF=\Nu+\Nd$):
\beq
\gamma(g)=\frac{11\CA-4\TF\NF}{6}\,.
\label{gammagdef}
\eeq
We also denote by $\hP_{ij}$ the ordinary function obtained
from the kernels $P_{ij}$ above by discarding the endpoint
contributions (i.e.~by turning plus distributions into ordinary
functions, and by ignoring contributions proportional to $\delta(1-z)$).
From eqs.~(\ref{Pqq})--(\ref{Pgg}) one can read off the quantities introduced
in eqs.~(\ref{AopBopCop}) and~(\ref{Oopform2}), since at this order:
\beq
\big(\Oop(z)\big)_{ij}=
\asotpi\Big(\APmat^{[0]}\Big)_{ij}\equiv
\asotpi\,P_{ij}(z)\,.
\label{OvsPLO}
\eeq
Thus:
\beqn
&&\tpioas\,A_q(z)=\CF\,\frac{1+z^2}{1-z}\,,\;\;\;\;
B_q=0\,,\;\;\;\;
C_{qq}(z)=0\,,
\label{qqABC}
\\
&&\tpioas\,C_{gq}(z)=\CF\,\frac{1+(1-z)^2}{z}\,,
\label{gqABC}
\\
&&\tpioas\,C_{qg}(z)=\TF\left(z^2+(1-z)^2\right)\,,
\label{qgABC}
\\
&&\tpioas\,\widetilde{A}_g(z)=2\CA\,z\,,\;\;\;\;
\tpioas\,\widetilde{B}_g=\gamma(g)\,,\;\;\;\;
\tpioas\,C_{gg}(z)=2\CA\left(\frac{1-z}{z}+z(1-z)\right).\phantom{aaaa}
\label{ggABC}
\eeqn
The case of a quark is straightforward: in view of eq.~(\ref{qqABC}),
by making the simplest choice\footnote{At this stage, this is not mandatory.
We shall comment further on this point (see sect.~\ref{sec:cons}).
\label{ft:BqIN}}
$B_q^{\Rinout}=0$, eq.~(\ref{bBoplg}) leads to:
\beqn
\overline{B}_q^\Rout&=&0\,,
\label{Bqgt}
\\
\overline{B}_q^\Rin&=&-\asotpi\int_0^1 dz\hP_{qq}(z)\stepf_{qq,z}^\Rin
\label{Bqlt}
\\*&\equiv&
-\asotpi\int_0^1 dz\,\half\left(\hP_{qq}(z)\stepf_{qq,z}^\Rin+
\hP_{gq}(z)\stepf_{gq,z}^\Rin\right)\,,
\label{Bqlt2}
\eeqn
with the form in eq.~(\ref{Bqlt2}) identical to that in eq.~(\ref{Bqlt}) 
thanks to the \mbox{$z\leftrightarrow 1-z$} symmetry of both the splitting
kernels and their respective integration limits (owing to eq.~(\ref{epflavLO})).
The corresponding quark Sudakov factor is obtained by inserting $\overline{B}_q^\Rin$ 
into eq.~(\ref{Sidef2}). The case of the gluon is slightly more involved. 
We use:
\beqn
\tpioas\,\tilde{b}_{gg}(z)&=&\frac{\CA}{1-z}+\frac{\CA}{z}-
\half\,\hP_{gg}(z)\equiv\CA\Big(2-z+z^2\Big)\,,
\label{tbgg}
\\
\tpioas\,\tilde{b}_{qg}(z)&=&-\half\hP_{qg}(z)\,,
\label{tbgq}
\eeqn
which indeed satisfies eq.~(\ref{BtBlocal}) given $\gamma(g)$
of eq.~(\ref{gammagdef}) (note that the sum over flavours includes
both quarks and antiquarks). With this, eqs.~(\ref{bBop2lg}) and~(\ref{ggABC})
lead to:
\beqn
\overline{B}_g^\Rout&=&-\asotpi\,\CA\int_0^1 dz\,z(1-z)\stepf_{gg,z}^\Rout
-\asotpi\,\half\sum_{q,\bq}\int_0^1 dz\,\hP_{qg}(z)\stepf_{qg,z}^\Rout\,,
\label{Bggt}
\\
\overline{B}_g^\Rin&=&-\asotpi\,\half
\int_0^1 dz\left(\hP_{gg}(z)\stepf_{gg,z}^\Rin+
\sum_{q,\bq}\hP_{qg}(z)\stepf_{qg,z}^\Rin\right)\,.
\label{Bglt}
\eeqn
The form of eq.~(\ref{Bglt}) stems from exploiting:
\beq
\int_0^1 dz\,\stepf_{gg,z}^\Rin
\left(-\frac{2\CA}{1-z}+\frac{\CA}{1-z}+\frac{\CA}{z}\right)=0\,,
\label{CAsimpl}
\eeq
which is due to the fact that (see eq.~(\ref{epflavLO})):
\beq
\epsilon_{gg}^{{\sss\rm L}}=\ep_g\,,\;\;\;\;
\epsilon_{gg}^{{\sss\rm U}}=\ep_g\;\;\;\;\Longrightarrow\;\;\;\;
\stepf_{gg,z}^\Rin=\stepf(\ep_g<z<1-\ep_g)\,.
\label{stepgg}
\eeq
If the range in $z$ defined by $\stepf_{gg,z}^\Rin$ were not symmetric 
under $z\leftrightarrow 1-z$, $\overline{B}_g^\Rin$ would still be well
defined, but eqs.~(\ref{bBop2lg}), (\ref{BtBgllocal}), and~(\ref{tbgg})
would not lead to a result solely expressed in terms of $\hP_{gg}$
for the part proportional to $\CA$. Finally, we point out that
eqs.~(\ref{Bqlt2}) and~(\ref{Bglt}), which enter the quark and gluon
Sudakov form factors (in the latter case, only when $\lambda=1$), 
respectively, have the usual form of the integrals of the splitting kernels 
over the inner region. The reader is encouraged to bear in mind that this 
is a consequence of several arbitrary choices, which we have outlined
in sect.~\ref{sec:ref}.

By using the results above, those of eqs.~(\ref{ZKop}) and~(\ref{WJop2}),
and the replacement of eq.~(\ref{lamrepl}), we obtain the following after 
some trivial algebra:
\beqn
\big(\Zop\left[F\right]\!\big)_q(x)&=&
\asotpi\int_0^1\frac{dz}{z}\,\stepf(z\ge x)
\left[\stepf_{qq,z}^\Rin\hP_{qq}(z)f_q\left(\frac{x}{z}\right)+
\stepf_{qg,z}^\Rin\hP_{qg}(z)f_g\left(\frac{x}{z}\right)\right],\phantom{aa}
\label{Zopqres}
\\
\big(\Wop\left[F\right]\!\big)_q(x)&=&
\asotpi\int_0^1 dz\,\Bigg\{\stepf_{qq,z}^\Rout\hP_{qq}(z)
\left[\frac{1}{z}\,f_q\left(\frac{x}{z}\right)\stepf(z\ge x)
-f_q(x)\right]
\nonumber\\*&&
\phantom{\asotpi\int_0^1 dz\,\stepf_\ep}
+\frac{1}{z}\,\stepf_{qg,z}^\Rout\hP_{qg}(z)
f_g\left(\frac{x}{z}\right)\stepf(z\ge x)\Bigg\}+
\lambda\overline{B}_q^\Rout f_q(x)\,,\phantom{aaa}
\label{Wopqres}
\eeqn
and:
\beqn
\big(\Zop\left[F\right]\!\big)_g(x)&=&
\asotpi\int_0^1 \frac{dz}{z}\,\stepf(z\ge x)
\left[\stepf_{gg,z}^\Rin\hP_{gg}(z)f_g\left(\frac{x}{z}\right)+
\sum_{q,\bq}\stepf_{gq,z}^\Rin\hP_{gq}(z)f_q\left(\frac{x}{z}\right)
\right],\phantom{aaaaa}
\label{Zopgres}
\\
\big(\Wop\left[F\right]\!\big)_g(x)&=&
\asotpi\int_0^1 dz\,\Bigg\{\stepf_{gg,z}^\Rout
\left[\frac{\hP_{gg}(z)}{z}f_g\left(\frac{x}{z}\right)\stepf(z\ge x)
-\frac{2\CA z}{1-z}\,f_g(x)\right]
\nonumber\\*&&\phantom{\asotpi\int_0^1 dz}
+\frac{\stepf(z\ge x)}{z}\sum_{q,\bq}\stepf_{gq,z}^\Rout
\hP_{gq}(z)f_q\left(\frac{x}{z}\right)\Bigg\}+
\lambda\overline{B}_g^\Rout f_g(x)\,.\phantom{aaa}
\label{Wopgres}
\eeqn
We note that, owing to eq.~(\ref{Bqgt}), the last term on the r.h.s.~of
eq.~(\ref{Wopqres}) is null, independent of the value of $\lambda$; the reader 
must bear in mind that this is a choice (see footnote~\ref{ft:BqIN}). If one 
chooses $\lambda=1$, eqs.~(\ref{tbgg}) and~(\ref{tbgq}) allow one to rewrite 
eq.~(\ref{Wopgres}) in the seemingly more familiar form:
\beqn
\!\!\!\!\big(\Wop\left[F\right]\!\big)_g(x)&=&
\asotpi\int_0^1 dz\,\Bigg\{\stepf_{gg,z}^\Rout\hP_{gg}(z)
\left[\frac{1}{z}f_g\left(\frac{x}{z}\right)\stepf(z\ge x)
-\half f_g(x)\right]
\nonumber\\*&&
\phantom{a}
+\sum_{q,\bq}\left(
\frac{1}{z}\,\stepf_{gq,z}^\Rout\hP_{gq}(z)f_q\left(\frac{x}{z}\right)
\stepf(z\ge x)
-\half\,\stepf_{qg,z}^\Rout\hP_{qg}(z)f_g\left(x\right)\right)\!\Bigg\}.
\phantom{aaa}
\label{Wopgreslameq1}
\eeqn
Again, here a simplification has been made thanks to the fact that 
the analogue of eq.~(\ref{CAsimpl}) holds with
\mbox{$\stepf_{gg,z}^\Rin\to\stepf_{gg,z}^\Rout$} there,
given eq.~(\ref{stepgg}). Moreover, we observe that this is also a direct
consequence of the fact that the subtraction term in eq.~(\ref{Wopgres})
is proportional to \mbox{$z/(1-z)$}, as opposed to \mbox{$1/(1-z)$} -- 
the definition of $\overline{\Bop}$ respects the convention for the plus 
prescription mentioned in footnote~\ref{ft:subtr}.

Equation~(\ref{Wopgreslameq1}) does not offer any
specific advantages w.r.t.~eq.~(\ref{Wopgres}). In addition to being
valid only when $\lambda=1$, it may seem to feature uncancelled 
divergences stemming from the second term in the integrand. In fact, this 
is not the case, as one can easily see by regularising the integral.
However, such a regularisation is not practical in the context
of numerical computations. A better alternative is to exploit the 
\mbox{$z\leftrightarrow 1-z$} symmetry of the $\hP_{gg}(z)$ and $\hP_{qg}(z)$ 
kernels and eq.~(\ref{stepgg}) (as well as its analogue for the
$g\to q\bq$ branching), and to obtain a manifestly-finite integral 
by means of either of the formal replacements:
\beq
\half f_g\left(x\right)\;\longrightarrow\;
\stepf\!\left(z\ge\half\right)f_g\left(x\right)\,,
\;\;\;\;\;\;\;\;
\half f_g\left(x\right)\;\longrightarrow\;
z\,f_g\left(x\right)\,,
\eeq
in the second and fourth terms of the integrand.

\subsection{Results on backward evolution\label{sec:LOres}}
We present here some results obtained within the leading-order
framework outlined above. Since our objective is to illustrate issues
raised in previous sections, rather than to perform realistic
phenomenology, we consider two cases of a universal,
flavour-independent cutoff $\epLij=\epUij=\ep$. The first is relatively 
large and scale-independent, $\ep=0.1$, which serves to highlight the
differences between the various NEP formulations.  The second is slightly
more realistic from a parton-shower MC viewpoint, being scale dependent
and defined as $\ep=\;\mbox{(2 GeV)}/q$, where $q$ is the mass scale relevant
to the current computation (e.g.~in the Sudakov factor of eq.~(\ref{Sidef2}),
$q=\sqrt{\kappa^2}$). For the leading-order PDFs we adopt the CT18LO set of 
ref.~\cite{Yan:2022pzl}; the argument of $\as$ is taken to be a
mass-scale squared and, in keeping with ref.~\cite{Yan:2022pzl},
we have $\as(m_{\sss Z}^2)=0.135$.
%%%%%%%%%%%%%%%%%%%%%%%%%%%%%%%%%%%%%%%%%%%%%%%%%%%%%%%%%%%%%%%%%%%%%%%%%%%
\begin{figure}[htbp]
  \begin{center}
    \vspace*{-40mm}
    \includegraphics[scale=0.7]{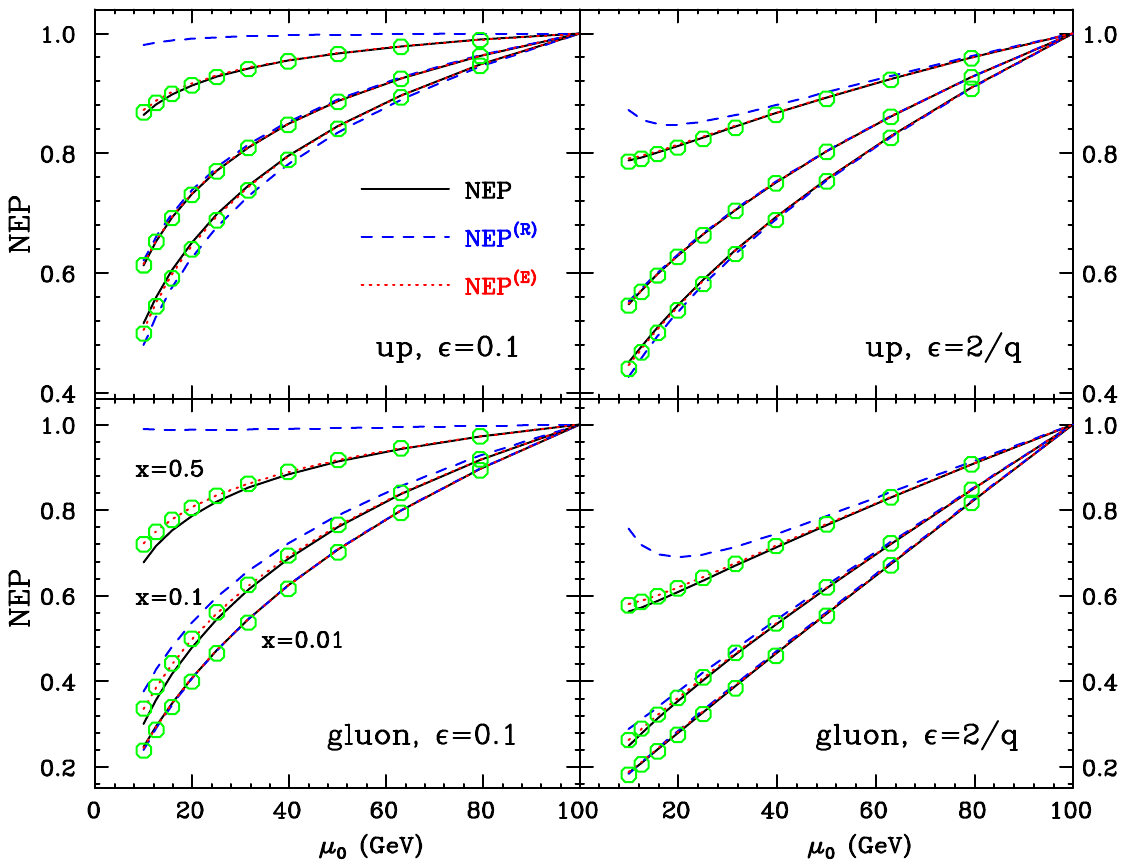}\vspace*{-40mm}
\caption{\label{fig:NEPgu}
  Non-emission probability (NEP) for backward evolution of up quarks
  and gluons with $\mu=100$ GeV,  according to NEP~(\ref{BnepJ})
  (black, solid), $\NEPbwR{}$~(\ref{BEstepR}) (blue, dashed)  and 
  $\NEPbwE{}$~(\ref{BEstepE}) (red, dotted).  The three sets of curves 
  correspond to $x=0.01$ (lowest), 0.1, and 0.5 (highest). The open circles 
  (green) show the NEP computed with cutoff-dependent PDFs, to be discussed in
  sect.~\ref{sec:PDFcut}.
}
\end{center}
\end{figure}
%%%%%%%%%%%%%%%%%%%%%%%%%%%%%%%%%%%%%%%%%%%%%%%%%%%%%%%%%%%%%%%%%%%%%%%%%%%
Figure~\ref{fig:NEPgu} shows the resulting true NEP~(\ref{BnepJ}) (black, 
solid) and the approximations $\NEPbwR{}$~(\ref{BEstepR}) (blue, dashed)  
and $\NEPbwE{}$~(\ref{BEstepE}) (red, dotted), for up-quarks and gluons
as a function of $\muz$, with $\mu=100$ GeV. In each panel,
the three sets of curves are for $x=0.01$, 0.1, and 0.5, from lowest
to highest, respectively.  One sees that, for the range of $\muz$
shown, $\NEPbwE{}$ is closer to the true NEP than $\NEPbwR{}$, as
could be anticipated from eqs.~(\ref{diffNEPR}) and~(\ref{diffNEPE3}).
$\NEPbwR{}$ becomes a poorer approximation with increasing $x$ (because
there $\Wop\left[F\right]$ tends to be large and negative), eventually 
possibly becoming non-monotonic and/or greater than unity at high $x$
(the latter e.g.~in the case of the down quark, which is not shown in the
figure).

\begin{figure}
  \begin{center}
\includegraphics[scale=0.4]{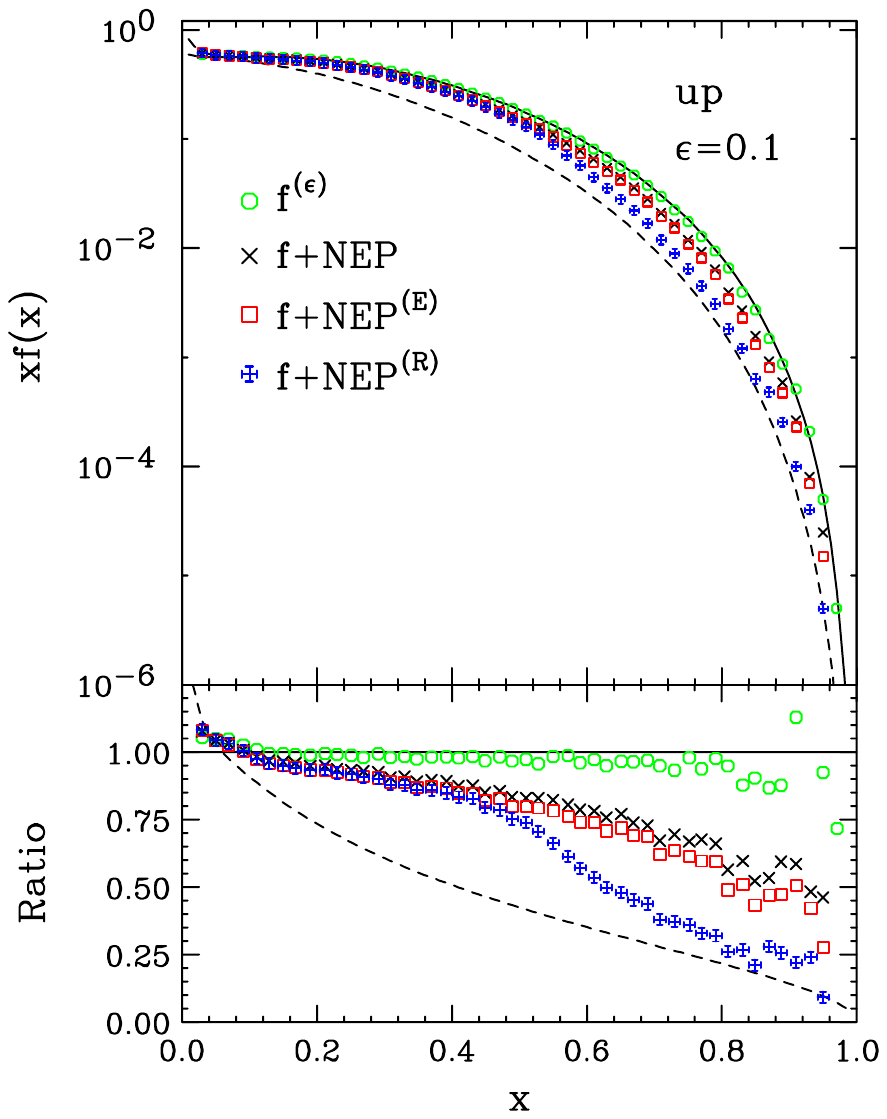}\hspace*{-49mm}
\includegraphics[scale=0.4]{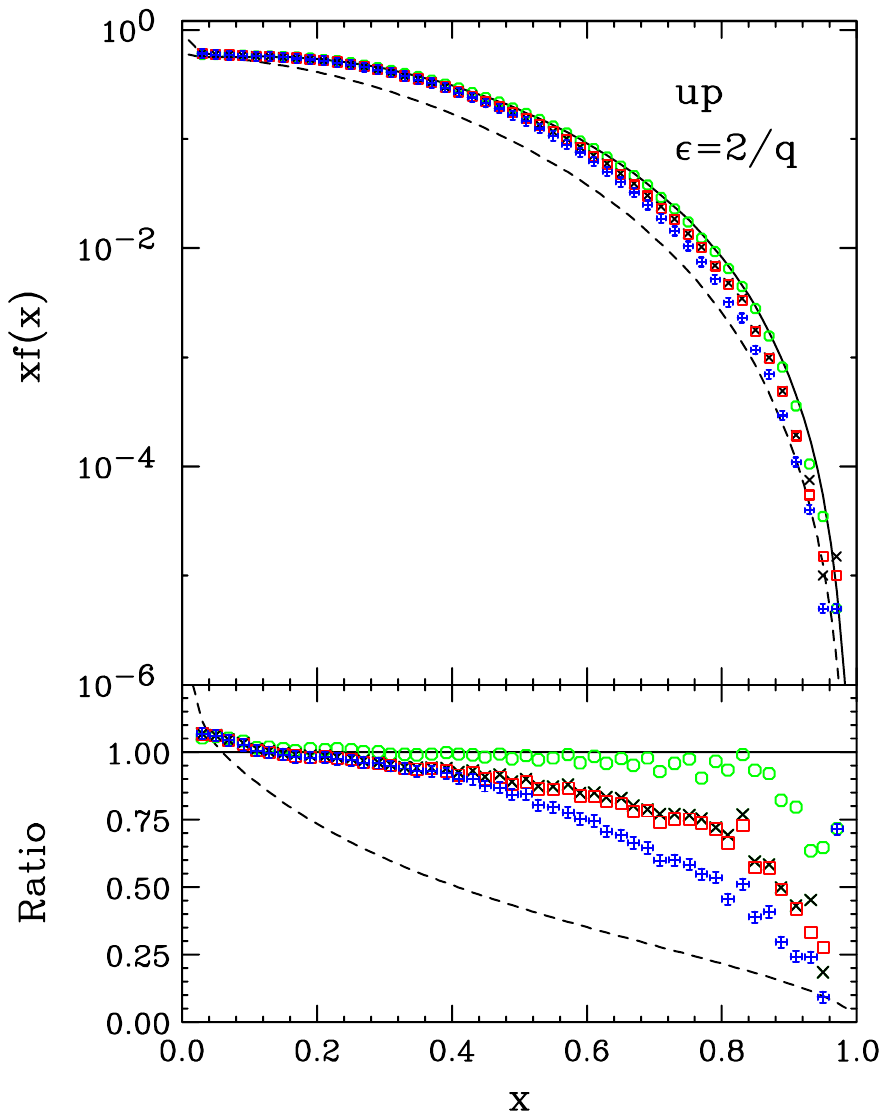}\vspace*{-10mm}
\includegraphics[scale=0.4]{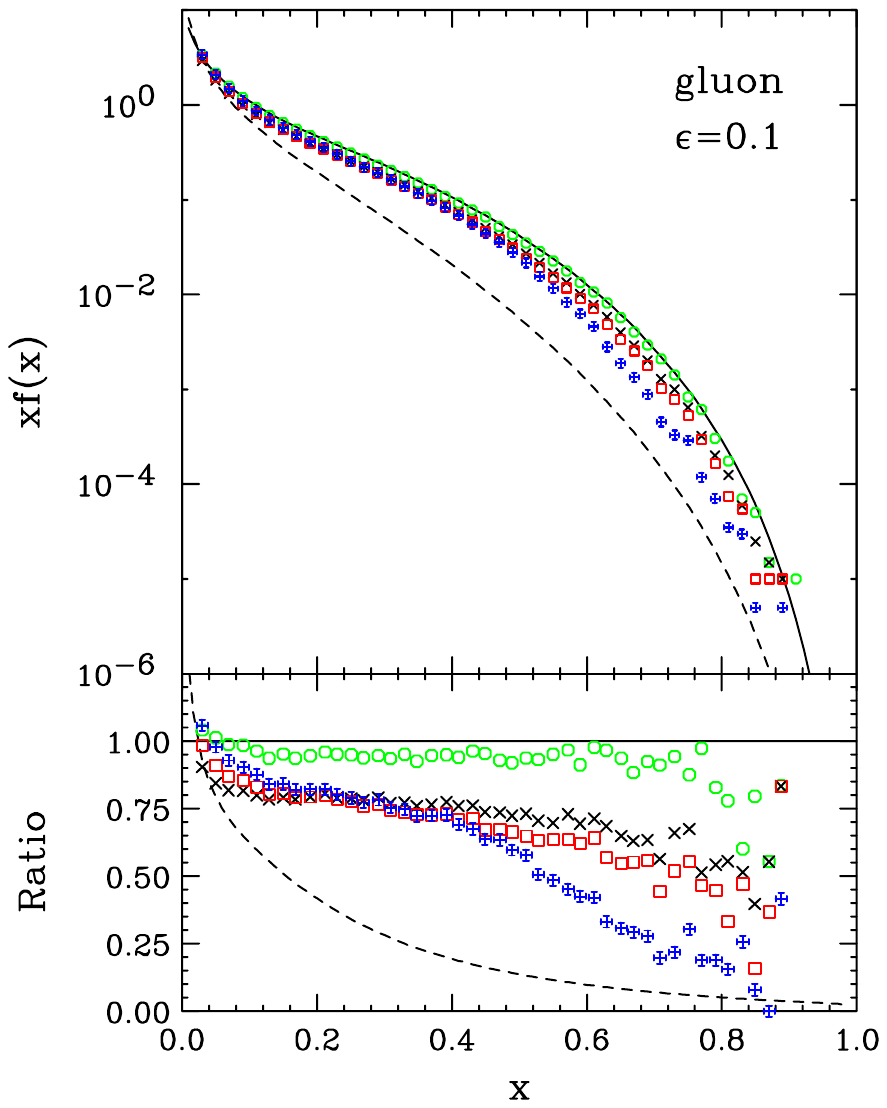}\hspace*{-49mm}
\includegraphics[scale=0.4]{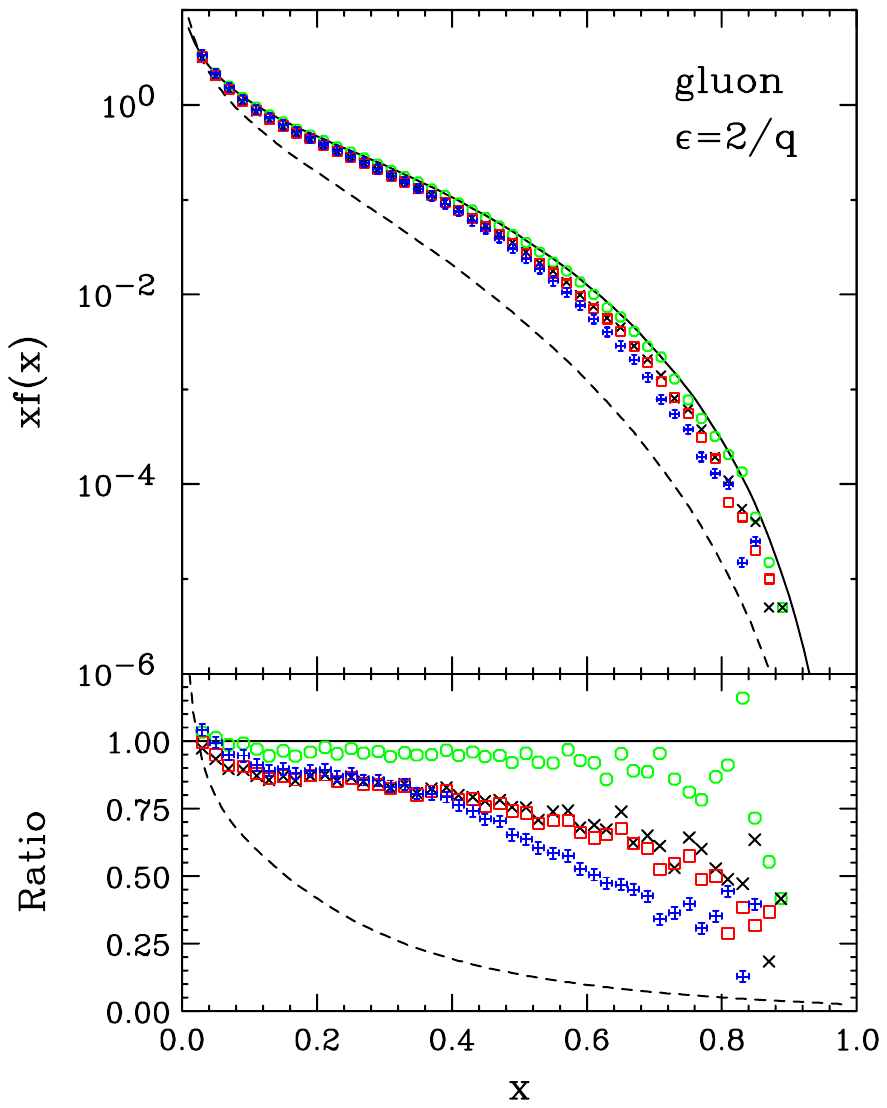}
\caption{\label{fig:NEPall}%
  Data points show up-quark and gluon PDFs at 10 GeV after MC backward
  evolution from 1 TeV, guided by the CT18LO PDFs using the NEP 
  (\ref{BnepJ}) (black crosses), and the approximation $\NEPbwR{}$
  (\ref{BEstepR}) (blue vertical crosses) or $\NEPbwE{}$ (\ref{BEstepE}) 
  (red boxes). Solid curves show the cutoff-independent PDFs at 10 GeV.  
  Also shown (dashed) are the cutoff-independent PDFs at
  the starting scale of 1 TeV, to illustrate the amount of
  evolution. As in fig.~\ref{fig:NEPgu}, open circles (green)
  show results obtained with cutoff-dependent
  PDFs, to be discussed in sect.~\ref{sec:PDFcut}. 
}
\end{center}
\end{figure}

\begin{figure}
  \begin{center}
\includegraphics[scale=0.4]{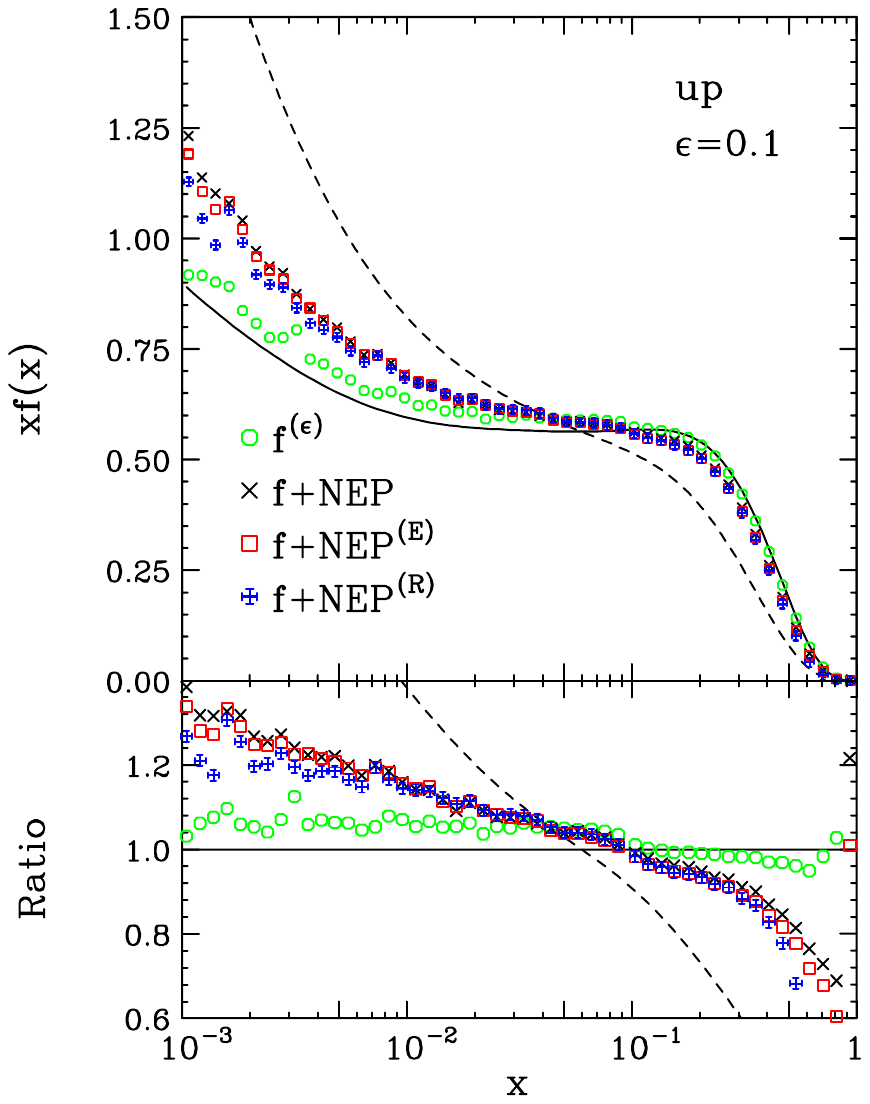}\hspace*{-49mm}
\includegraphics[scale=0.4]{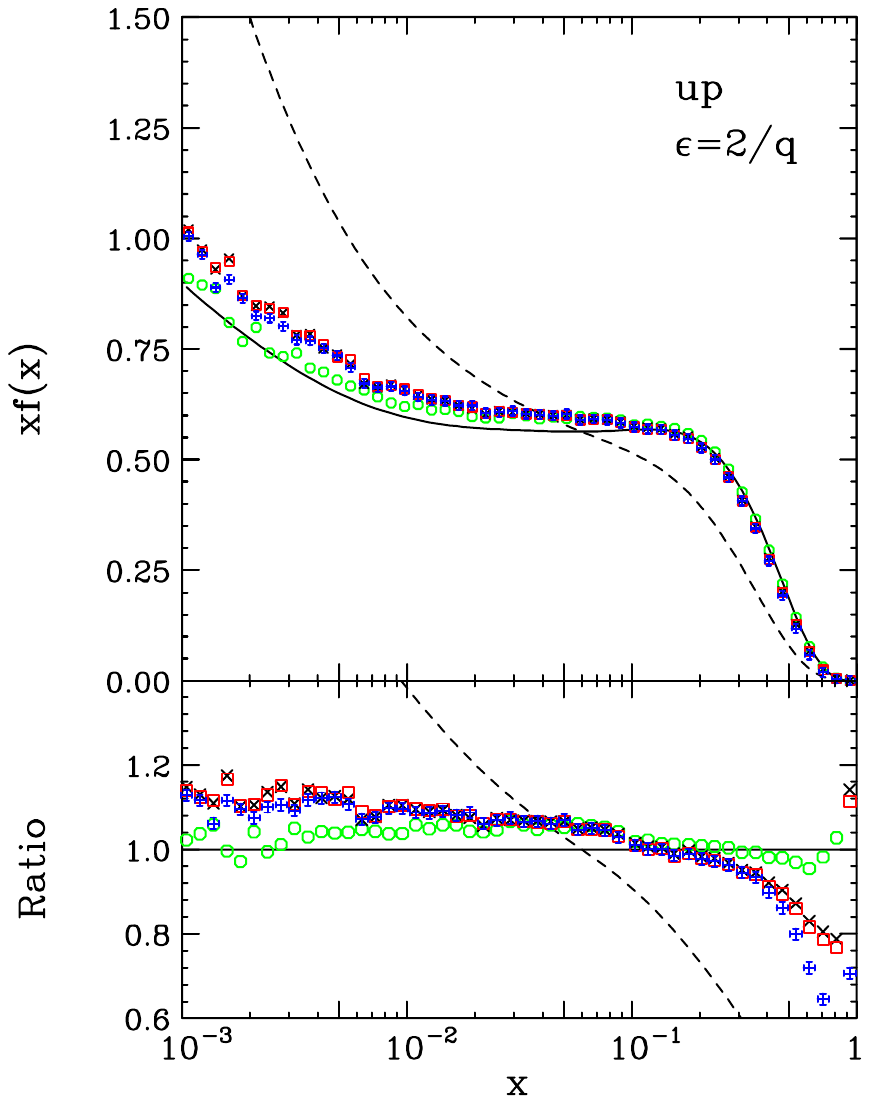}\vspace*{-10mm}
\includegraphics[scale=0.4]{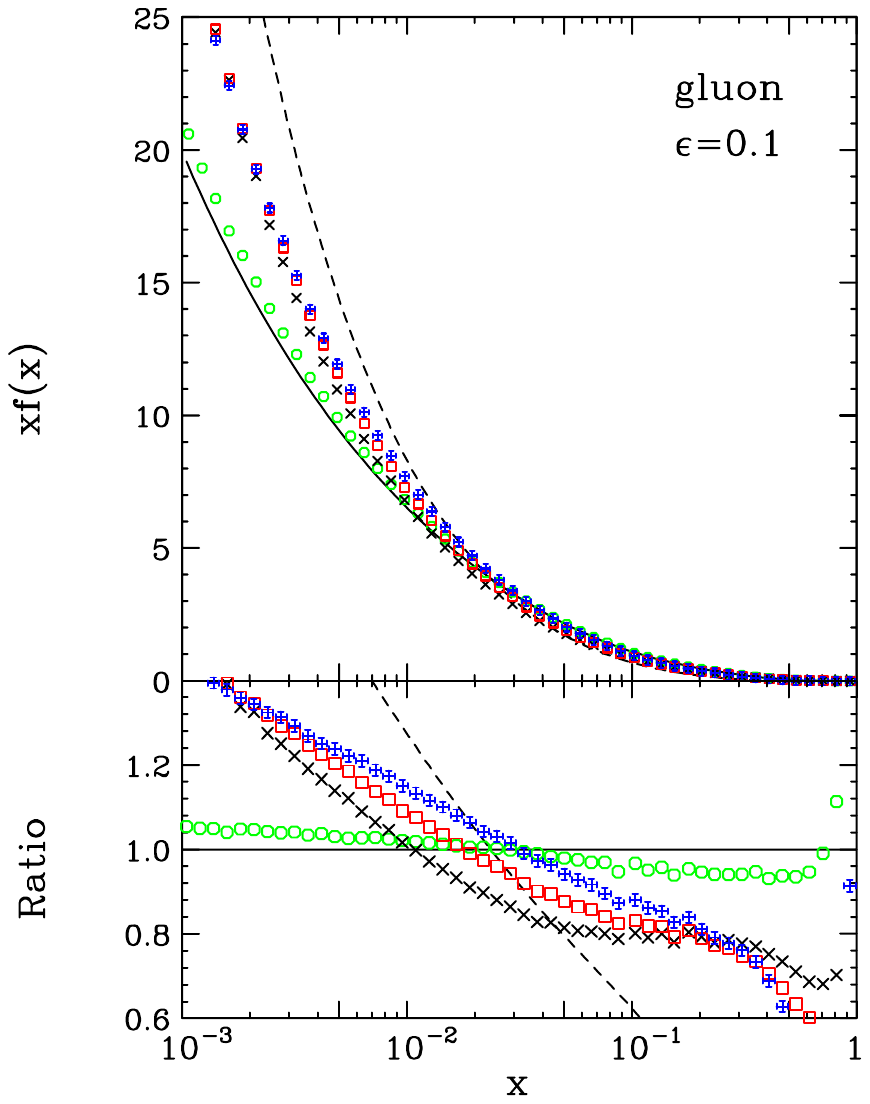}\hspace*{-49mm}
\includegraphics[scale=0.4]{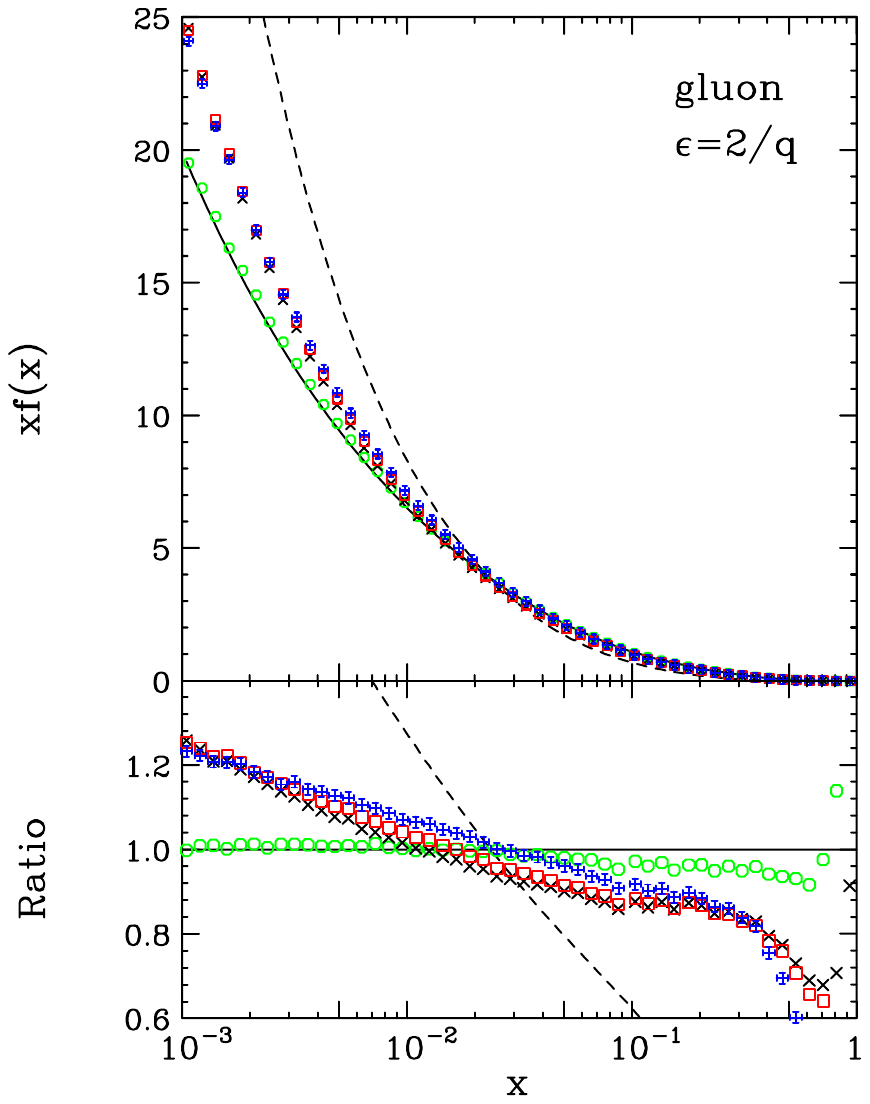}
\caption{\label{fig:NEPlog}%
  As in fig.~\ref{fig:NEPall}, but with a logarithmic $x$ scale.
  }
\end{center}
\end{figure}

Figures~\ref{fig:NEPall} and \ref{fig:NEPlog} show results of MC
backward evolution from 1 TeV to 10 GeV using the NEP~(\ref{BnepJ})
(black crosses), and the approximation $\NEPbwR{}$~(\ref{BEstepR}) (blue
vertical crosses) or $\NEPbwE{}$~(\ref{BEstepE}) (red boxes). Here $10^7$ 
unweighted MC events were generated starting at $\mu=1$ TeV, with a 
probability  distribution of momentum fraction $x$ and flavour $i$
\beq
\frac{dP_i}{dx} = x f_i(x,\mut)\,,
\eeq
using the momentum sum rule
\beq
\sum_j
\int_0^1 dx\,x f_j(x,\mut) =1
\eeq
as normalization.  Following the selection of the next branching
scale $\mu_0$ according to the relevant NEP, the momentum fraction
$x^\prime$ and flavour $j$ of the branching parent was chosen according to
the distribution
\beq
\frac{dP_j}{dx^\prime} = \frac 1{x^\prime}\stepf_{ij,x/x^\prime}^\Rin
\hP_{ij}(x/x^\prime)f_j(x^\prime,\muzt)
\Bigg/\sum_k\int_x^1\frac{dz}z\,\stepf_{ik,z}^\Rin
\hP_{ik}(z)f_k(x/z,\muzt)\,.
\label{dPjodx}
\eeq
Note that this implies that only resolvable emissions were generated;
although this is the standard practice, it is not necessarily what
the various NEPs employed here would dictate -- more details on this
point are given in app.~\ref{sec:PDFsol} (see in particular
eqs.~(\ref{probi1}) and~(\ref{proby10})).  

For computational speed, backward evolution was discretized on a
$(500,70)$-node grid in $(x,\mu)$, i.e.~evolution was restricted to
hopping between nodes, in order to avoid slow two-dimensional
interpolation. The procedure was iterated until the next branching
scale fell below 10 GeV. The $x$-grid was logarithmic from $10^{-6}$
to 0.1 (250 nodes) and linear above 0.1 (250 nodes); the $\mu$-grid
was logarithmic from 10 GeV to 1 TeV (35 nodes per decade). This
provided sufficient precision for comparative purposes. MC results
were then binned linearly or logarithmically in $x$, as required for
fig.~\ref{fig:NEPall} or \ref{fig:NEPlog}, respectively.

In the case of $\NEPbwR{}$, we have seen that it may be
non-monotonic or larger than one, in which case the solution of
eq.~(\ref{BEstepR}) was chosen larger than or equal to the value
of $\muz$ for which $\NEPbwR{}$ has its minimum.  This may
account for a part of the large discrepancies between the results of using
$\NEPbwR{}$ and  the true NEP or $\NEPbwE{}$ at high $x$.

Generally speaking, all versions of the NEP perform poorly in
reproducing the backward evolution of the PDFs, especially outside the
intermediate region $0.01<x<0.1$; the true NEP performs best at higher $x$.  
However, the fundamental problem remains that the PDF evolution generated by 
the MC results from the accumulation of recoils against resolved emissions,
whereas the actual evolution results from both resolved and unresolved
emissions\footnote{For an early prescription to account for unresolved 
emissions in an average manner, see ref.~\cite{Bengtsson:1986gz}.
\label{ft:py6}}.

One possible approach that avoids this problem, while introducing
others, is to guide the backward MC with cutoff-dependent PDFs that
are generated by resolved emissions alone.  We consider this
approach in detail in the following section.

%%%%%%%%%%%%%%%%%%%%%%%%%%%%%%%%%%%%%%%%%%%%%%%%%%%%%%%%%%%%%%%%%%%%%%%%%%%
%%%%%%%%%%%%%%%%%%%%%%%%%%%%%%%%%%%%%%%%%%%%%%%%%%%%%%%%%%%%%%%%%%%%%%%%%%%

%%%%%%%%%%%%%%%%%%%%%%%%%%%%%%%%%%%%%%%%%%%%%%%%%%%%%%%%%%%%%%%%%%%%%%%%%%%
%%%%%%%%%%%%%%%%%%%%%%%%%%%%%%%%%%%%%%%%%%%%%%%%%%%%%%%%%%%%%%%%%%%%%%%%%%%

\section{PDF evolution with cutoff\label{sec:PDFcut}}
In sect.~\ref{sec:gen} we have shown that the correct form of the
backward initial-state radiation NEP is that of eq.~(\ref{BnepJ}).
However, since the NEP is associated with the resolution-dependent
\mbox{(non-)} emission of resolvable partons, there is an inconsistency in
using it to generate backward shower evolution guided by PDFs that
obey the resolution-independent evolution equations (\ref{evolmu2}).
Barring some {\em ad hoc} reweighting procedure, this results in 
discrepancies between the guiding PDFs and those generated by 
backward evolution, which we have illustrated in the previous section.

An alternative approach is suggested by eq.~(\ref{NEPWeq0}).
Namely, one defines a new type of PDFs, which we denote by $F^{(\ep)}$,
that obey the following evolution equation\footnote{This possibility
was pointed out but not explored in ref.~\cite{Marchesini:1987cf}.}:
\beq
\frac{\partial F^{(\ep)}(x)}{\partial\log\mut}=\Oop^\Rin\otimes_x F^{(\ep)}\,.
\label{evolmu2ep}
\eeq
As the notation suggests, such PDFs depend on the cutoffs 
\mbox{$\ep=\{\epLij,\epUij\}_{ij}$}.
However, in view of eqs.~(\ref{evolmu4}) and~(\ref{OopgtW}),
and of the characteristics of $\Wop[F]$ (see in particular
eqs.~(\ref{WJop2}) and~(\ref{limBgt})), we expect that $F^{(\ep)}$
and $F$ will differ, {\em in the resolved region}, only by terms 
suppressed by some powers of the cutoffs; conversely, in the unresolved
region the differences between the two are in general logarithmic
in the cutoffs.

\subsection{Flavour and momentum conservation\label{sec:cons}}
One interesting question that immediately emerges in the case
of cutoff-dependent evolution, eq.~(\ref{evolmu2ep}), is whether
flavour and momentum are conserved. By working again at the LO, 
we obtain for the integrated non-singlet contribution of quark flavour $q$:
\beq
\frac{\partial}{\partial\log\mut}\int_0^1 dx
\Big(f_q^{(\ep)}(x)-f_{\bq}^{(\ep)}(x)\Big)=
\asotpi\int_0^1 dz\,\big(\Oop^\Rin(z)\big)_{qq}
\int_0^1 dy\Big(f_q^{(\ep)}(y)-f_{\bq}^{(\ep)}(y)\Big)\,,
\label{flavc}
\eeq
having used the identity:
\beq
\int_0^1 dx\,g\otimes_x h=\int_0^1 dz\,g(z)\int_0^1 dy\,h(y)\,.
\eeq
With eqs.~(\ref{Ooplt}) and~(\ref{qqABC}) we obtain:
\beqn
\int_0^1 dz\,\big(\Oop^\Rin(z)\big)_{qq}&=&
\int_0^1 dz\,A_q(z)\stepf_{qq,z}^\Rin+\overline{B}_q^\Rin
\label{momcq1}
\\*&=&
0
\label{momcq2}
\\*&=&
B_q^\Rin\,.
\label{momcq3}
\eeqn
The result of eq.~(\ref{momcq2}) stems from eq.~(\ref{Bqlt}),
whereas that of eq.~(\ref{momcq3}) is what we would have obtained
if we had not chosen $B_q^\Rin=0$ (see eq.~(\ref{bBoplg})).

This gives us the opportunity to discuss a general property of the
Sudakov definition in the context of MC-compatible PDF evolution
equations. In particular, one observes that this definition is to
a certain extent always arbitrary. In the standard case, such an
arbitrariness is associated with the choice of the resolved region --
in practice, with the choices of the cutoffs and of the functional
dependence upon them of the borders of the resolved region.
In the case of cutoff-dependent evolution, in addition to the
above there is the freedom associated with the choice of the 
parameters $\Bop^{\Rinout}$ (or $\widetilde{\Bop}^{\Rinout}$,
the two being related to each other by relationships such as
eq.~(\ref{bvstbint})), given $\Bop$ and the constraints
of eq.~(\ref{Bopaux}) (or given $\widetilde{\Bop}$ and the constraints
of eq.~(\ref{wBopaux})). Once these choices have been made, what
is exponentiated ($\overline{\Bop}^\Rin$) is determined unambiguously
by eq.~(\ref{bBoplg}) or eq.~(\ref{bBop2lg}). In the case of a
quark, $B_q=0$, and eq.~(\ref{Bopaux}) implies that $B_q^\Rin$ can be
set equal to any function of the cutoffs that vanishes with them.
Thus, while eq.~(\ref{momcq2}) shows that the cutoff-dependent
evolution conserves flavour given the choice of eq.~(\ref{Bqlt}),
eq.~(\ref{momcq3}) can be seen as constraining $B_q^\Rin$ by {\em imposing}
that flavour be conserved. In other words: the additional freedom of
the cutoff-dependent evolution w.r.t.~the standard one discussed
above, namely that associated with terms suppressed by powers of the
cutoffs, can be exploited to impose physical conditions (such as
flavour conservation) that, at variance with the case of standard
evolution, may not necessarily emerge in a natural manner.

Turning to the case of momentum conservation\footnote{The issue of momentum
conservation in PDF evolution with a cutoff, in that case due to a modified
argument of $\as$, was considered in ref.~\cite{Skrzypek:2007zz}.}, we write:
\beqn
&&\frac{\partial}{\partial\log\mut}\int_0^1 dx\,x\sum_i f_i^{(\ep)}(x)=
\label{momc}
\\*&&\phantom{aaaa}
\asotpi\int_0^1 dz\,z\Big(\big(\Oop^\Rin(z)\big)_{gg}+
\sum_{q,\bq}\big(\Oop^\Rin(z)\big)_{qg}\Big)
\int_0^1 dy\,y\,f_g^{(\ep)}(y)
\nonumber
\\*&&\phantom{aa}\!
+\asotpi\sum_{q}\int_0^1 dz\,z
\Big(\big(\Oop^\Rin(z)\big)_{qq}+\big(\Oop^\Rin(z)\big)_{gq}\Big)
\int_0^1 dy\,y\Big(f_q^{(\ep)}(y)+f_{\bq}^{(\ep)}(y)\Big),
\nonumber
\eeqn
having used the identity:
\beq
\int_0^1 dx\,x\,g\otimes_x h=\int_0^1 dz\,z\,g(z)\int_0^1 dy\,y\,h(y)\,.
\eeq
Let us start by considering the integral over $x$ in the third line
of eq.~(\ref{momc}). By proceeding as was done for the manipulations
of the flavour-conservation case, we obtain:
\beqn
&&\int_0^1 dz\,z
\Big(\big(\Oop^\Rin(z)\big)_{qq}+\big(\Oop^\Rin(z)\big)_{gq}\Big)=
\label{intmomc1}
\\*&&\phantom{aaaa}
\int_0^1 dz\,z\left[A_q(z)\stepf_{qq,z}^\Rin+
C_{gq}(z)\stepf_{gq,z}^\Rin\right]+B_q^\Rin
-\int_0^1 dz\,A_q(z)\stepf_{qq,z}^\Rin=
B_q^\Rin\,,
\nonumber
\eeqn
where the rightmost side follows from the direct computation of the
integrals that appear in the central expression. Since $B_q^\Rin=0$ as we have
discussed above, the integral on the l.h.s.~of eq.~(\ref{intmomc1})
is thus equal to zero. Turning to the integral over $z$ in the second 
line of eq.~(\ref{momc}), we have:
\beqn
&&\int_0^1 dz\,z\Big(\big(\Oop^\Rin(z)\big)_{gg}+
\sum_{q,\bq}\big(\Oop^\Rin(z)\big)_{qg}(z)\Big)=
\label{intmomc2}
\\*&&\phantom{aa}
\int_0^1 dz\,z\left[\left(\frac{\widetilde{A}_g(z)}{1-z}
+C_{gg}(z)\right)\stepf_{gg,z}^\Rin+
\sum_{q,\bq}C_{qg}(z)\stepf_{qg,z}^\Rin\right]+\widetilde{B}_g^\Rin
-\int_0^1 dz\,\frac{\widetilde{A}_g(1)}{1-z}\stepf_{gg,z}^\Rin\,.
\nonumber
\eeqn
By using eqs.~(\ref{ggABC}), (\ref{tbgg}), and~(\ref{tbgq}) (the latter
two in eq.~(\ref{BtBgllocal})), one sees that the integral on the l.h.s.~of
eq.~(\ref{intmomc2}) is equal to zero. Combined with the null result 
in eq.~(\ref{intmomc1}), this is the analogue of eq.~(\ref{momcq2}),
and concludes the proof that the cutoff-dependent evolution conserves
the momentum. Conversely, we can proceed by analogy with eq.~(\ref{momcq3}),
and impose the l.h.s.~of eq.~(\ref{intmomc2}) to be equal to zero
in order to determine $\widetilde{B}_g^\Rin$.  If we then adopt the
local form (\ref{BtBgllocal}):
\beq
\widetilde{B}^{\Rin}_g = \int_0^1 dz\left[\tilde{b}_{gg}(z)
  \stepf_{gg,z}^\Rin+\sum_{q,\bq}\tilde{b}_{qg}(z)
  \stepf_{qg,z}^\Rin\right],
\eeq
we obtain
\beqn
\tpioas\,\tilde{b}_{gg}(z)&=&2\CA\,z\Big(2-z+z^2\Big)\,,
\label{tbgg2}
\\
\tpioas\,\tilde{b}_{qg}(z)&=&-z\hP_{qg}(z)\,,
\label{tbgq2}
\eeqn
leading via eq.~(\ref{Sidef2}) to the following expression for the
integrand in the exponent of the gluon Sudakov form factor:
\beq
\overline{B}_g^\Rin=-\asotpi
\int_0^1 dz\,z\left(\hP_{gg}(z)\stepf_{gg,z}^\Rin+
\sum_{q,\bq}\hP_{qg}(z)\stepf_{qg,z}^\Rin\right)\,.
\label{Bglt2}
\eeq
Although different from  eq.~(\ref{Bglt}), this expression is equally
valid, since the two integrands differ by a function that 
integrates to zero. In an analogous manner, from eq.~(\ref{intmomc1}) we 
would obtain for the quark
\beq
\overline{B}_q^\Rin=-\asotpi\int_0^1 dz\,z\left(\hP_{qq}(z)\stepf_{qq,z}^\Rin+
\hP_{gq}(z)\stepf_{gq,z}^\Rin\right)\,,
\label{Bqlt3}
\eeq
which again coincides with eq.~(\ref{Bqlt2}), in spite of having a
different integrand. We point out that the strict equality of
the results for $\overline{B}_g^\Rin$ stemming from eqs.~(\ref{Bglt2}) 
and~(\ref{Bglt}), and of those for $\overline{B}_q^\Rin$ from
eqs.~(\ref{Bqlt3}) and~(\ref{Bqlt2}), relies among other things on
the symmetry properties of the $\stepf_{ij,z}^\Rin$ functions.
On the other hand, the integrands of eqs.~(\ref{Bglt2}) and~(\ref{Bqlt3}), 
at variance with those of eqs.~(\ref{Bglt}) and~(\ref{Bqlt2}), do not have 
a $z\to 0$ singularity when $\ep\to 0$. This implies that they lead to
finite quantities also when completely removing the constraints enforced by
the lower cutoffs; such quantities can then be employed to define Sudakov 
factors that differ from those used thus far by terms suppressed by powers 
of the cutoff\footnote{For examples of Sudakov form factors in a 
different context, whose definitions do differ from one another by 
cutoff-suppressed terms, see e.g.~app.~A of
ref.~\cite{Frederix:2020trv}.}.

\subsection{Results on cutoff-dependent PDFs\label{sec:PDFcutres}}
%%%%%%%%%%%%%%%%%%%%%%%%%%%%%%%%%%%%%%%%%%%%%%%%%%%%%%%%%%%%%%%%%%%%%%%%%%%
\begin{figure}
  \begin{center}
\includegraphics[scale=0.4]{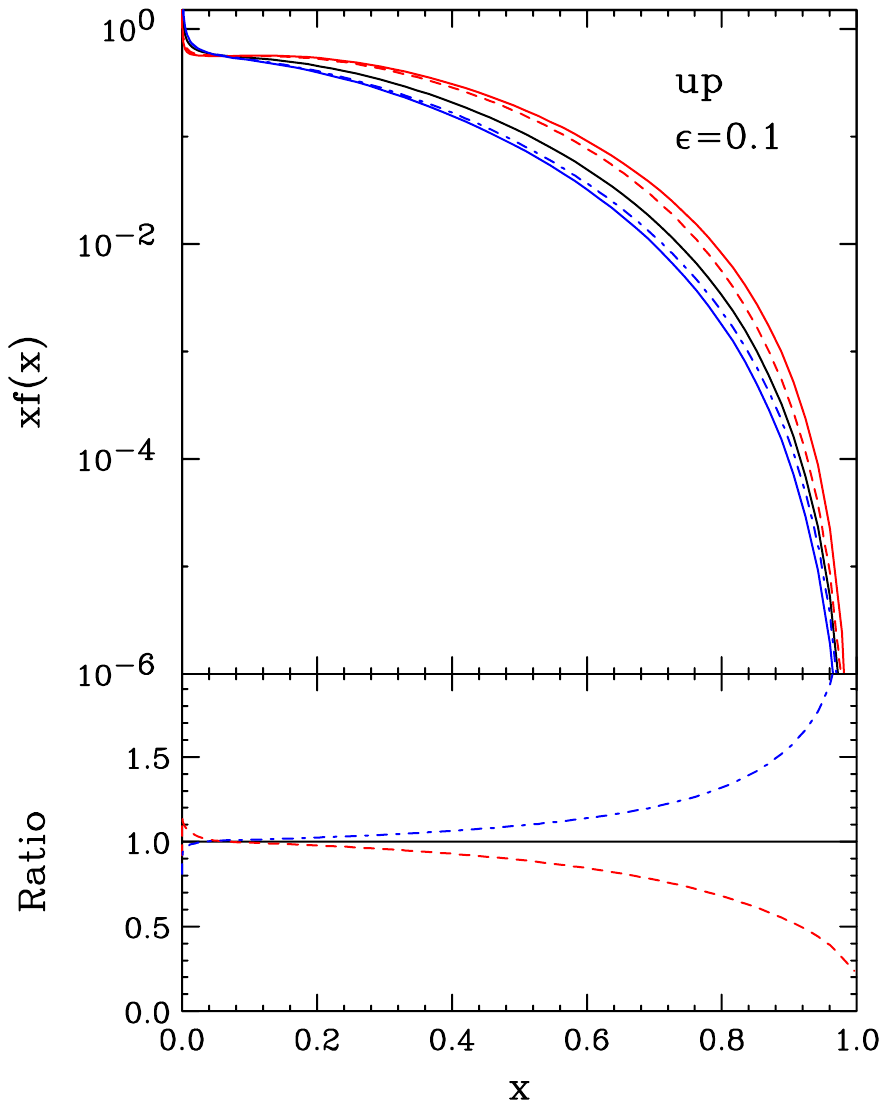}\hspace*{-49mm}
\includegraphics[scale=0.4]{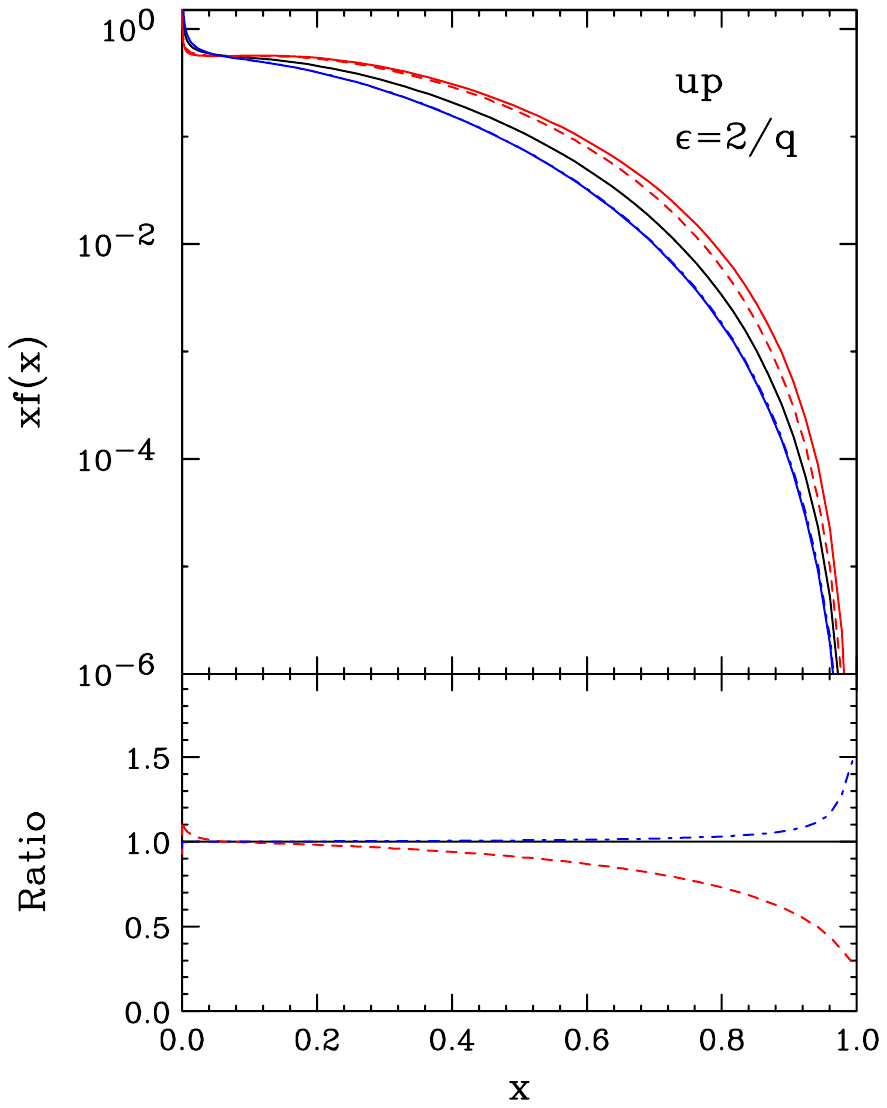}\vspace*{-10mm}
\includegraphics[scale=0.4]{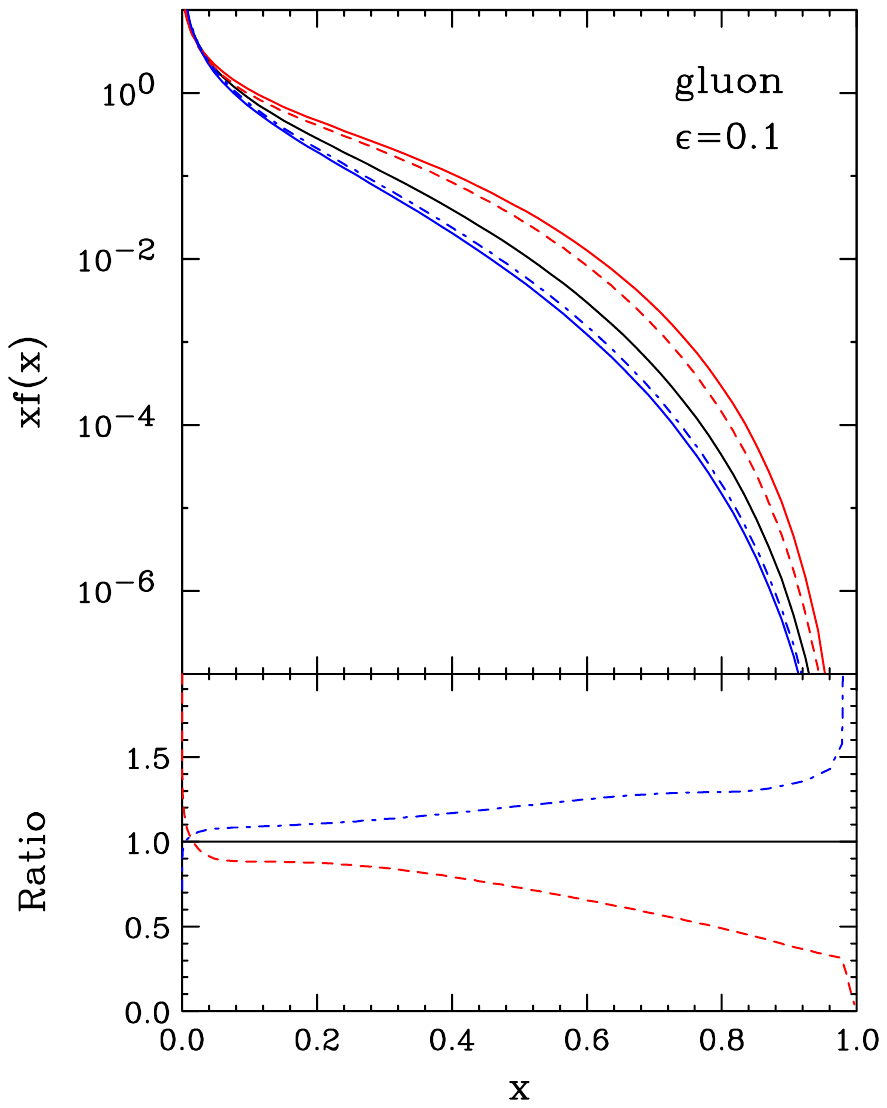}\hspace*{-49mm}
\includegraphics[scale=0.4]{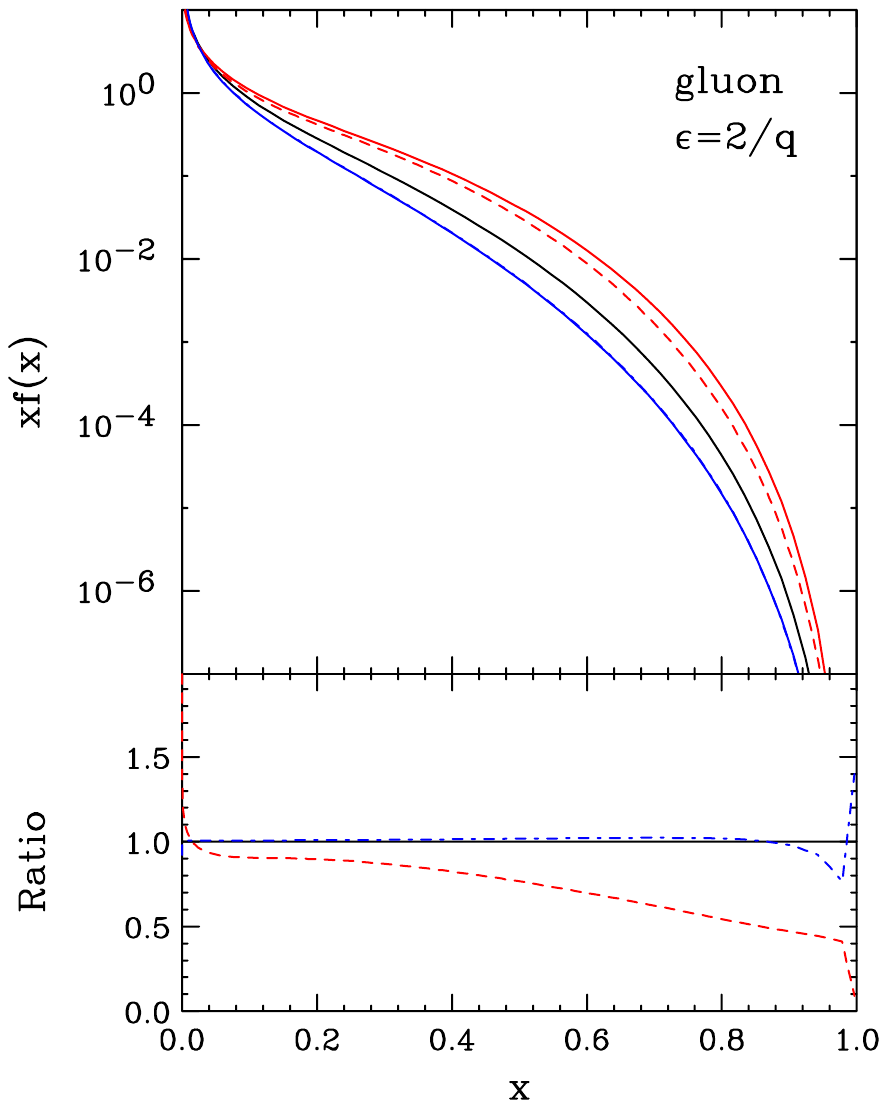}
\caption{\label{fig:pdfcut}%
 Cutoff-dependent PDFs evolved from the CT18LO set at 100 GeV
 backwards to 10 GeV (red, dashed) and forwards to 1 TeV
 (blue, dot-dashed), compared to cutoff-independent PDFs (solid) at 10 GeV
 (red), 100 GeV (black) and 1 TeV  (blue).  The ratio plots show the
 cutoff-dependent PDFs relative to the cutoff-independent ones at the 
 same scale.
  }
\end{center}
\end{figure}
%%%%%%%%%%%%%%%%%%%%%%%%%%%%%%%%%%%%%%%%%%%%%%%%%%%%%%%%%%%%%%%%%%%%%%%%%%%

Figure~\ref{fig:pdfcut} shows examples of cutoff-dependent PDFs
corresponding to the two cutoff choices discussed in
sect.~\ref{sec:LOres}.  Starting from the CT18LO set at scale
$\mu=100$ GeV, the PDFs were evolved forwards to 1 TeV and backwards
to 10 GeV using eq.~(\ref{evolmu2ep}) in place of (\ref{evolmu2}),
with the flavour- and momentum-conserving formulation described
above.\footnote{The effects of cutoffs on (forward) PDF evolution are
  also considered in ref.~\cite{Mendizabal:2023mel}.}

As expected, the cutoff-dependent PDFs generally evolve more slowly with
increasing scale than the true PDFs, thus being generally softer below
the starting scale and harder above it.  The relative differences grow
with increasing $x$ and are largest in the unresolved region
$x>1-\ep$, where the PDFs are however very small.  The scale-dependent
cutoff is smaller than the scale-independent one over most of the
evolution range and naturally leads to PDFs closer to the 
cutoff-independent ones.

Non-emission probabilities for backward evolution guided by the
cutoff-dependent PDFs, chosen to coincide with the CT18LO set at 10 GeV,
are shown by green circles in fig.~\ref{fig:NEPgu} as a function of
$\muz$ with $\mu=100$~GeV.  Since by construction
$\Wop\left[F^{(\ep)}\right]\equiv 0$, all of the expressions 
for the NEP are now equivalent, by virtue of
eq.~(\ref{NEPWeq0}). Empirically, the NEP for cutoff-dependent PDFs
appears closest to  $\NEPbwE{}$ computed from cutoff-independent PDFs.

Results of backward MC evolution guided by the cutoff-dependent PDFs are 
also shown by green circles in figs.~\ref{fig:NEPall} and \ref{fig:NEPlog}. 
There, in contrast to fig.~\ref{fig:pdfcut} but analogously to 
fig.~\ref{fig:NEPgu}, the cutoff-dependent PDFs
were chosen to coincide with the CT18LO set at 10 GeV and evolved
upwards to 1 TeV, where they were used as the starting distributions
for the backward MC.  In this way, the backward-generated MC
distributions at 10 GeV should agree with the CT18LO set.
Compared to the results using cutoff-independent PDFs, agreement
is indeed greatly improved at all $x$ values.  The small residual systematic
discrepancies are most likely due to accumulated errors from our
discretization of the backward evolution.

\section{Cutoff-dependent cross sections\label{sec:xsecs}}
The cutoff-dependent PDFs emerging from eq.~(\ref{evolmu2ep}) imply
that short-distance cross sections must be cutoff-dependent too,
in order for the l.h.s.~of the factorisation formula to be cutoff 
independent\footnote{Up to terms one perturbative order higher than 
those included in the computation of the short distance cross sections.}. 
We shall assume in what follows that the cutoff dependence of the PDFs 
is solely due to their evolution. This implies that, at the scale chosen 
as the starting point for PDF evolution, the initial conditions must be 
cutoff independent; this is not mandatory, but doing otherwise would 
require some modeling assumptions for the initial conditions.
In order to determine the cutoff-dependent terms of the cross section, 
we consider the generic factorisation formula for one incoming leg,
starting from the cutoff-independent case:
\beq
\sigma=F^{\rm T}\star\hat{\Sigma}\equiv
\sum_i\int_0^1 dx\,f_i(x)\,\hsig_i(x)\,.
\label{fact0}
\eeq
Here, we have denoted by $\hat{\Sigma}$ the column vector that
collects all of the (subtracted) short-distance cross sections
\mbox{$\hsig_i\equiv(\hat{\Sigma})_i$}, whereas $\sigma$ is the
hadron-level cross section that results from the sum over all of
the partonic processes in eq.~(\ref{fact0}). The RGE invariance
of $\sigma$ under factorisation-scale variation is:
\beq
0=\frac{\partial\sigma}{\partial\log\mut}=
\frac{\partial F^{\rm T}}{\partial\log\mut}\star\hat{\Sigma}
+F^{\rm T}\star\frac{\partial\hat{\Sigma}}{\partial\log\mut}=
\left(\Oop\otimes F\right)^{\rm T}\star\hat{\Sigma}
+F^{\rm T}\star\frac{\partial\hat{\Sigma}}{\partial\log\mut}\,.
\label{RGE1}
\eeq
It is a matter of algebra to show that, for any functions $g$, $h$, and $l$,
the following identity holds:
\beq
\big(g\otimes h\big)\star l=
g\star\big(h\star l)=h\star\big(g\star l)\,.
\eeq
Equation~(\ref{RGE1}) then implies:
\beq
0=F^{\rm T}\star\frac{\partial\hat{\Sigma}}{\partial\log\mut}
+F^{\rm T}\star\left(\Oop^{\rm T}\star\hat{\Sigma}\right).
\eeq
This equation must be true for any PDFs, and therefore:
\beq
\frac{\partial\hat{\Sigma}}{\partial\log\mut}=
-\Oop^{\rm T}\star\hat{\Sigma}
\;\;\;\;\Longleftrightarrow\;\;\;\;
\frac{\partial\hsig_i(x)}{\partial\log\mut}=
-\sum_j\int_0^1 dy\,\big(\Oop(y)\big)_{ji}\,\hsig_j(xy)\,.
\label{hsigsol}
\eeq
By writing the perturbative expansion of the short-distance
cross sections as follows:
\beq
\hat{\Sigma}=\hat{\Sigma}^{[0]}+\asotpi\,\hat{\Sigma}^{[1]}+\ldots
\eeq
where all terms $\hat{\Sigma}^{[i]}$ include a factor $\as^b$,
with $b$ a process-dependent constant (e.g.~$b=0$ and $b=2$ for
dilepton and Higgs production, respectively), and by working with
LO kernels where eq.~(\ref{OvsPLO}) holds, eq.~(\ref{hsigsol}) implies:
\beq
\hat{\Sigma}^{[1]}=
-\log\frac{\mut}{\qzt}\,
\left(\tpioas\Oop^{{\rm T}}\star\hat{\Sigma}^{[0]}\right)+C
\equiv
-\log\frac{\mut}{\qzt}\,
\left(\APmat^{[0]{\rm T}}\star\hat{\Sigma}^{[0]}\right)+C\,,
\label{hsigsol0}
\eeq
with $\qz$ an arbitrary reference scale, and $C$ a column vector of
$\mu$-independent integration constants. The determination of $C$ can 
be done by means of an explicit cross section calculation. For example, 
it can be read from the FKS formalism~\cite{Frixione:1995ms,Frixione:1997np}, 
where a term in the same form as the leftmost one on the r.h.s.~of 
eq.~(\ref{hsigsol0}) is contained in the so-called $(n+1)$-body degenerate 
contributions.

The derivation above can be repeated verbatim for the cutoff-dependent
PDFs and short-distance cross sections. Denoting the latter by 
$\hat{\Sigma}^{(\ep)}$, owing to eq.~(\ref{evolmu2ep}) the analogue of 
eq.~(\ref{hsigsol0}) reads as follows:
\beq
\hat{\Sigma}^{(\ep)[1]}=
-\log\frac{\mut}{\qzt}\,
\left(\tpioas\big(\Oop^\Rin\big)^{\rm T}\star
\hat{\Sigma}^{(\ep)[0]}\right)+C^{(\ep)}\,.
\label{hsigsol1}
\eeq
Under our assumptions concerning the cutoff dependence discussed at
the beginning of this section, we may now set:
\beq
\hat{\Sigma}^{(\ep)[0]}=\hat{\Sigma}^{[0]}\,.
\eeq
Furthermore, by choosing $\qz$ to coincide with the starting scale of the
PDF evolution, at $\mu=\qz$ we must have:
\beq
\hat{\Sigma}^{(\ep)[1]}=\hat{\Sigma}^{[1]}\;\;\;\;\Longrightarrow\;\;\;\;
C^{(\ep)}=C\,,
\eeq
and therefore, for a generic scale value:
\beq
\hat{\Sigma}^{(\ep)[1]}=\hat{\Sigma}^{[1]}
+\log\frac{\mut}{\qzt}\,\left[\tpioas
\left(\Oop-\Oop^\Rin\right)^{{\rm T}}\star\hat{\Sigma}^{[0]}\right]
\equiv\hat{\Sigma}^{[1]}
+\log\frac{\mut}{\qzt}\,
\left(\tpioas\big(\Oop^\Rout\big)^{\rm T}\star\hat{\Sigma}^{[0]}\right)\,.
\label{Ccorr}
\eeq
Equation~(\ref{Ccorr}) allows one to obtain the sought cutoff-dependent
short-distance cross sections given the cutoff-independent ones.
The rightmost term on the r.h.s.~of eq.~(\ref{Ccorr}) is, as expected,
suppressed by powers of the cutoff; we shall call it the cutoff
correction.

We note that by iteration of this procedure one can obtain the
cutoff correction to any perturbative order, in terms of contributions
of lower orders to the short-distance cross section and
the cutoff-dependent and cutoff-independent evolution kernels.

\subsection{Results for cross sections\label{sec:res}}
\subsubsection{Drell-Yan process\label{sec:DY}}
As a first illustration of the use of cutoff-dependent PDFs with a
cutoff-corrected short-distance cross section, we consider the 
photon-induced $\ord(\aem^2)$ and $\ord(\aem^2\as)$ 
contributions to the cross section for lepton pair production as a
function of pair invariant mass $\mll$ at fixed hadronic collision
energy $\sqrt s$.

\begin{figure}
  \begin{center}
\includegraphics[scale=0.4]{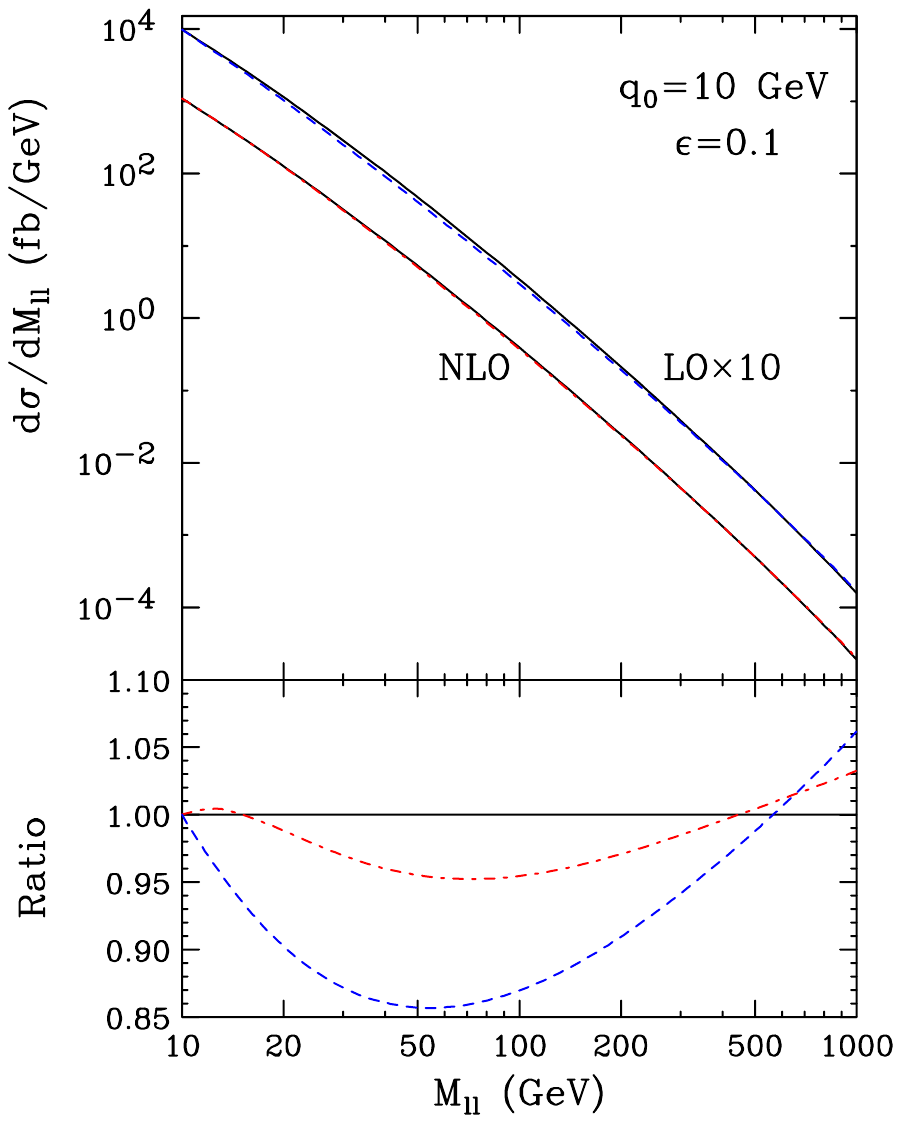}\hspace*{-49mm}
\includegraphics[scale=0.4]{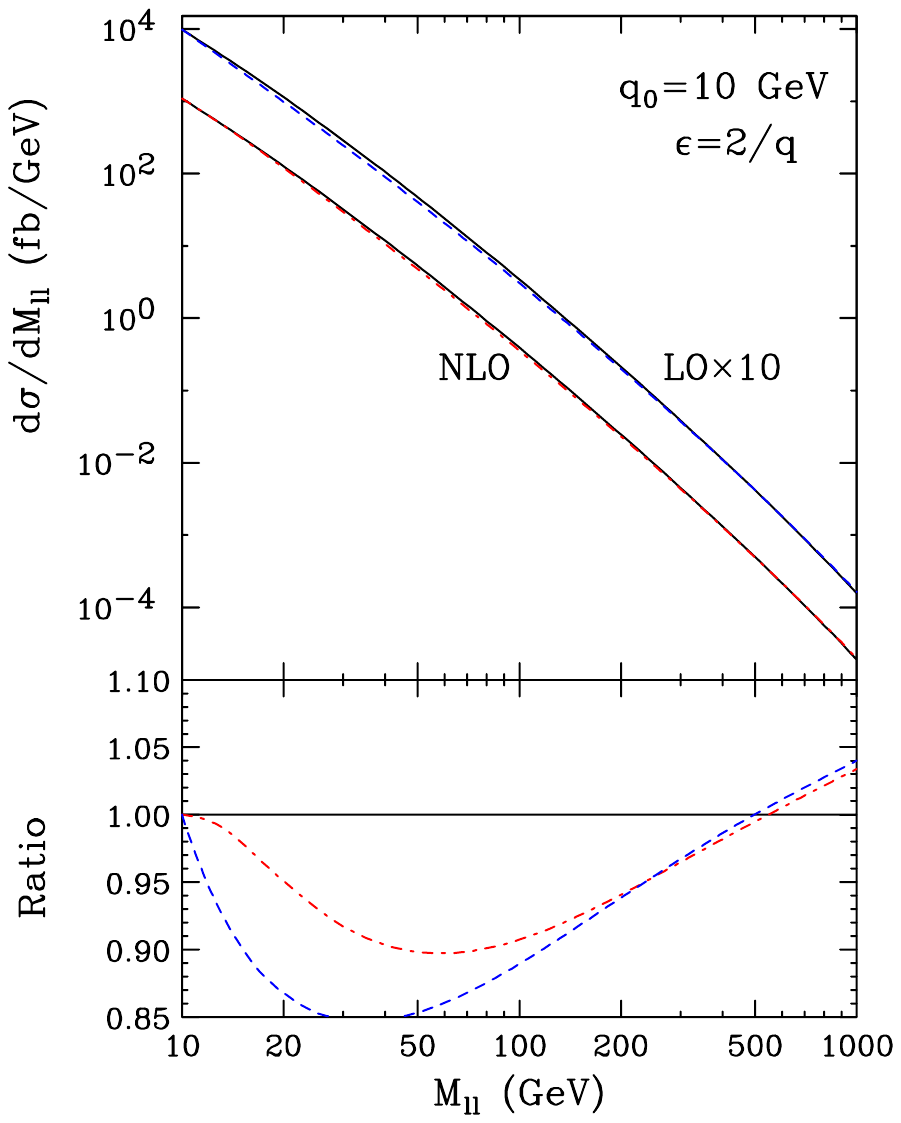}\vspace*{-10mm}
\includegraphics[scale=0.4]{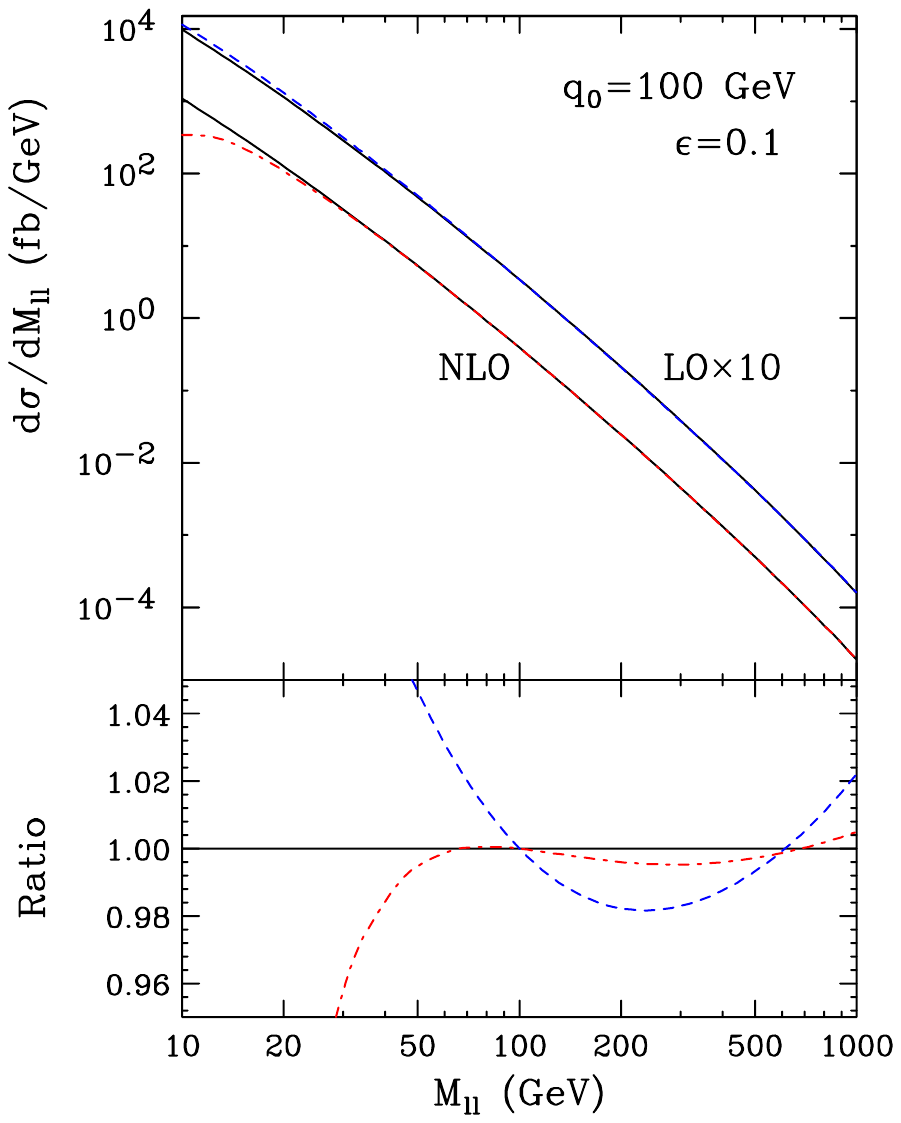}\hspace*{-49mm}
\includegraphics[scale=0.4]{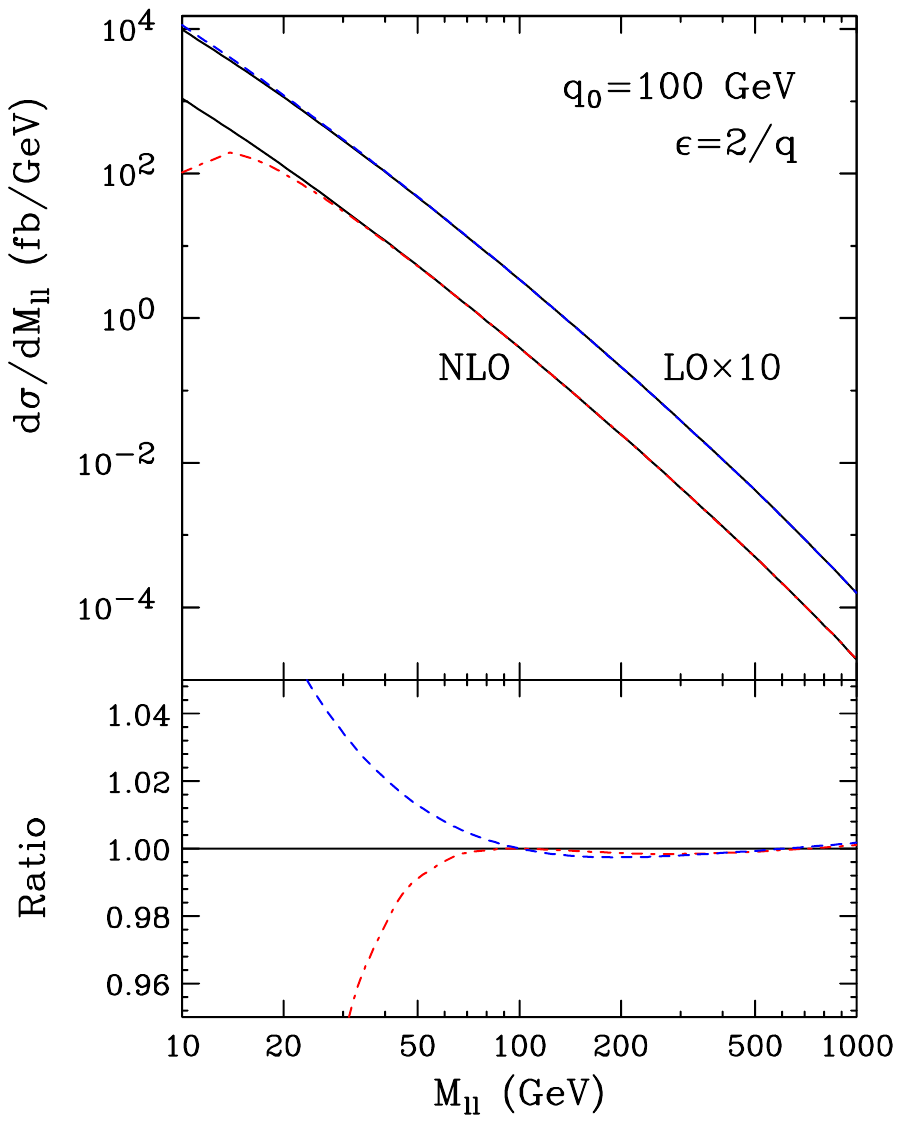}
\caption{\label{fig:dycs}%
  Drell-Yan cross section at $\sqrt s=13$ TeV (photon-induced contribution 
  only), calculated using cutoff-dependent PDFs at leading order (blue,
  dashed) and next-to-leading order (red, dot-dashed), compared to
  corresponding results using cutoff-independent PDFs (solid).
   }
\end{center}
\end{figure}

Some results for $pp$ collisions at $\sqrt s=13$ TeV are shown in
fig.~\ref{fig:dycs}.  As in sect.~\ref{sec:LOres}, we use the CT18LO
leading-order PDFs, and we do so with both LO ($\ord(\aem^2)$) 
and NLO ($\ord(\aem^2+\aem^2\as)$) short-distance cross sections, for 
the latter of which we employ the $\overline{\rm MS}$ factorisation scheme.
We again consider two cases of a universal, flavour-independent cutoff
$\epLij=\epUij=\ep$: one relatively  large and scale-independent,
$\ep=0.1$,  the other scale-dependent, $\ep=\;\mbox{(2 GeV)}/q$, with 
$q$ the relevant mass scale.  The reference scale $q_0$, at which the
cutoff-dependent and cutoff-independent PDFs are identical, is set equal to
10 GeV in the upper plots and to 100 GeV in the lower ones.  The scale
for evaluation of the PDFs, $\as$, and NLO corrections is taken to be
$\mut=\mll^2$ throughout.

At leading order there is no cutoff correction to the short-distance
cross section and so the discrepancies between the cutoff-dependent
(blue, dashed) and cutoff-independent (black, solid) LO results simply
reflect those between the corresponding PDFs.  Since the cutoff-dependent
PDFs evolve more slowly, the resulting hadronic cross section initially 
falls below the true (i.e.~obtained with cutoff-independent PDFs)
LO value for $\mll>q_0$ but eventually rises above it 
at higher values of $\mll$ (higher $x$).  Correspondingly, for $q_0=100$ GeV, 
it lies above the true LO value when $\mll<q_0$.

At next-to-leading order, the cutoff correction comes into play and
reduces the discrepancy between the cutoff-dependent and true NLO
results.  The reduction is strong around the reference
scale $\mll\sim q_0$, but only modest above and very rapidly
deteriorating below $q_0$.  For the scale-dependent cutoff with a low
reference scale (the upper right plot), the effect of the cutoff
correction vanishes much more rapidly than that of the difference in
PDFs at high $\mll$.  This is because the relevant scale in the cutoff
correction is the local value $q=\mll$, whereas the difference in PDFs
results from the accumulation of cutoff effects over the whole range
from $q_0$ to $\mll$.

In summary, the comparison of the LO and NLO results shows that the
NLO cutoff correction of eq.~(\ref{Ccorr}) partly compensates for the 
differences between the cutoff-independent predictions and those one
would have obtained by employing cutoff-dependent PDFs without the inclusion of
such a correction in the short-distance cross sections. In general, the use 
of cutoff-dependent PDFs together with the correction (\ref{Ccorr})
gives results for the Drell-Yan cross section
that are relatively close to the cutoff-independent NLO predictions, provided 
the reference scale $q_0$ is close to, or not too far below, the dilepton
mass. We point out that, at this level of accuracy, a more systematic 
assessment of the compensation mechanism just mentioned would require
the definition of a proper NLO cutoff-dependent PDF set.
%%%%%%%%%%%%%%%%%%%%%%%%%%%%%%%%%%%%%%%%%%%%%%%%%%%%%%%%%%%%%%%%%%%%%%%%%%%
%%%%%%%%%%%%%%%%%%%%%%%%%%%%%%%%%%%%%%%%%%%%%%%%%%%%%%%%%%%%%%%%%%%%%%%%%%%

\subsubsection{Higgs boson production\label{sec:H}}
Since the Drell-Yan process is quark dominated, we consider as a
second example the gluon fusion contribution to Higgs boson 
hadroproduction as a function of the hadronic collision energy $\sqrt s$.
%%%%%%%%%%%%%%%%%%%%%%%%%%%%%%%%%%%%%%%%%%%%%%%%%%%%%%%%%%%%%%%%%%%%%%%%%%%
\begin{figure}
  \begin{center}
\includegraphics[scale=0.4]{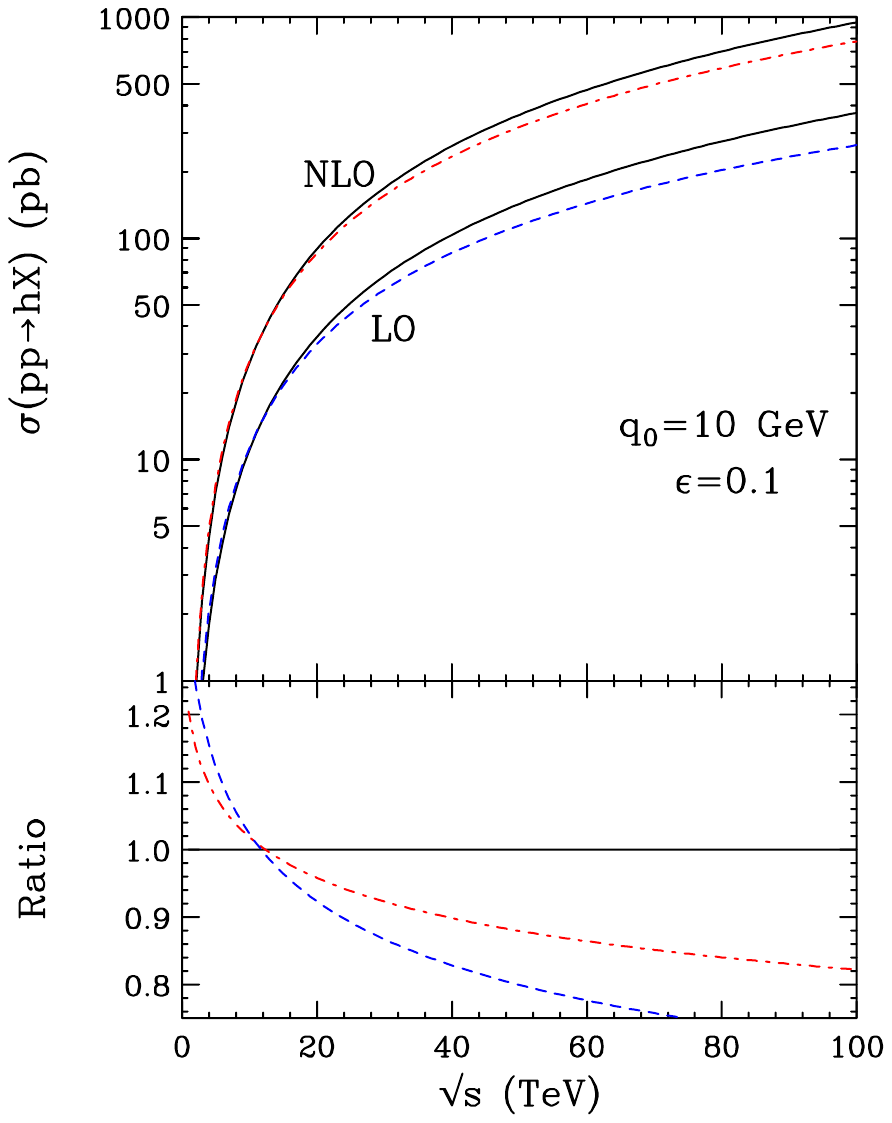}\hspace*{-49mm}
\includegraphics[scale=0.4]{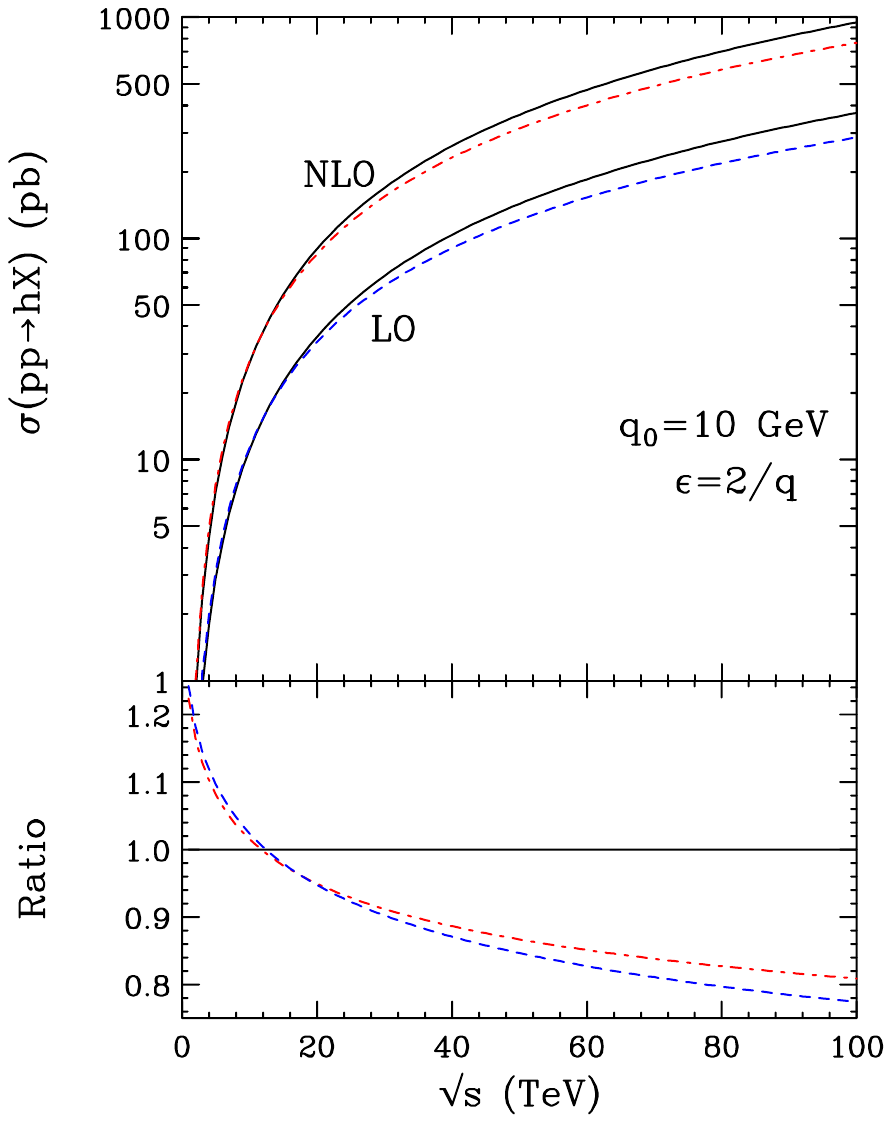}\vspace*{-10mm}
\includegraphics[scale=0.4]{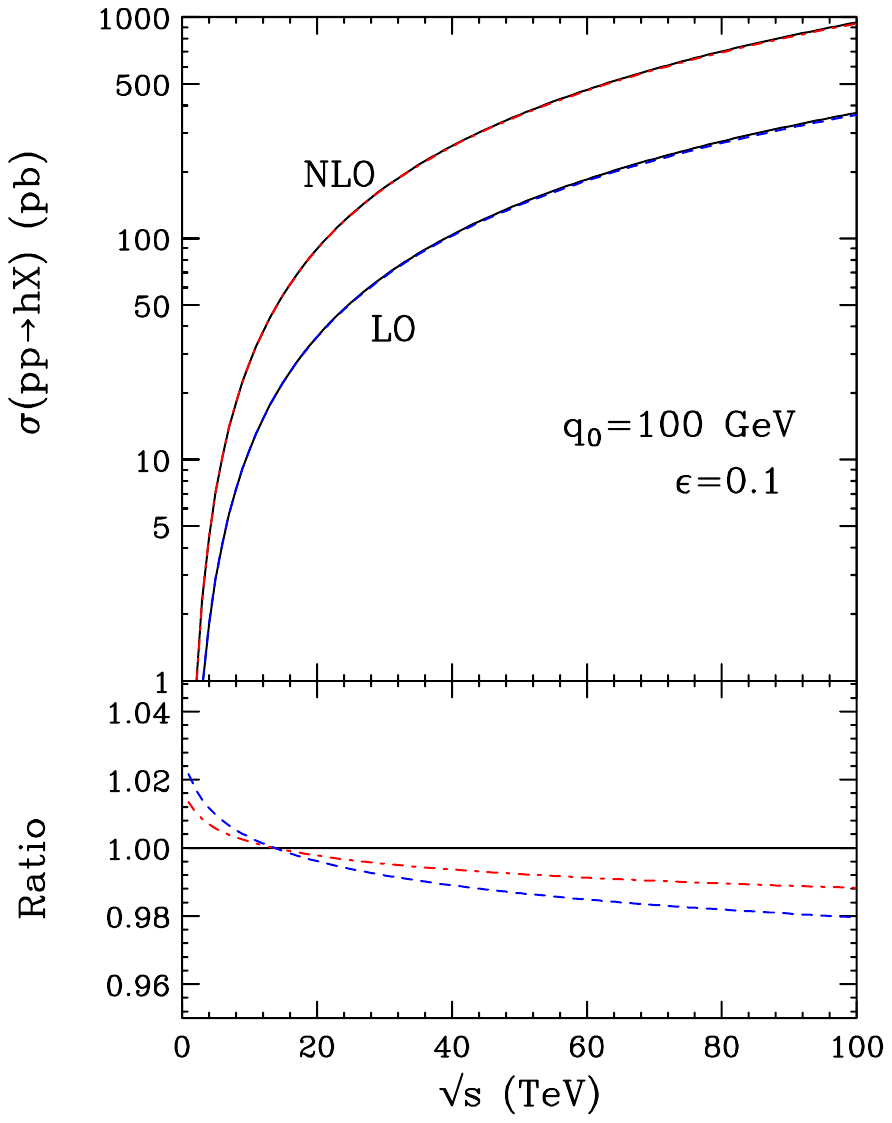}\hspace*{-49mm}
\includegraphics[scale=0.4]{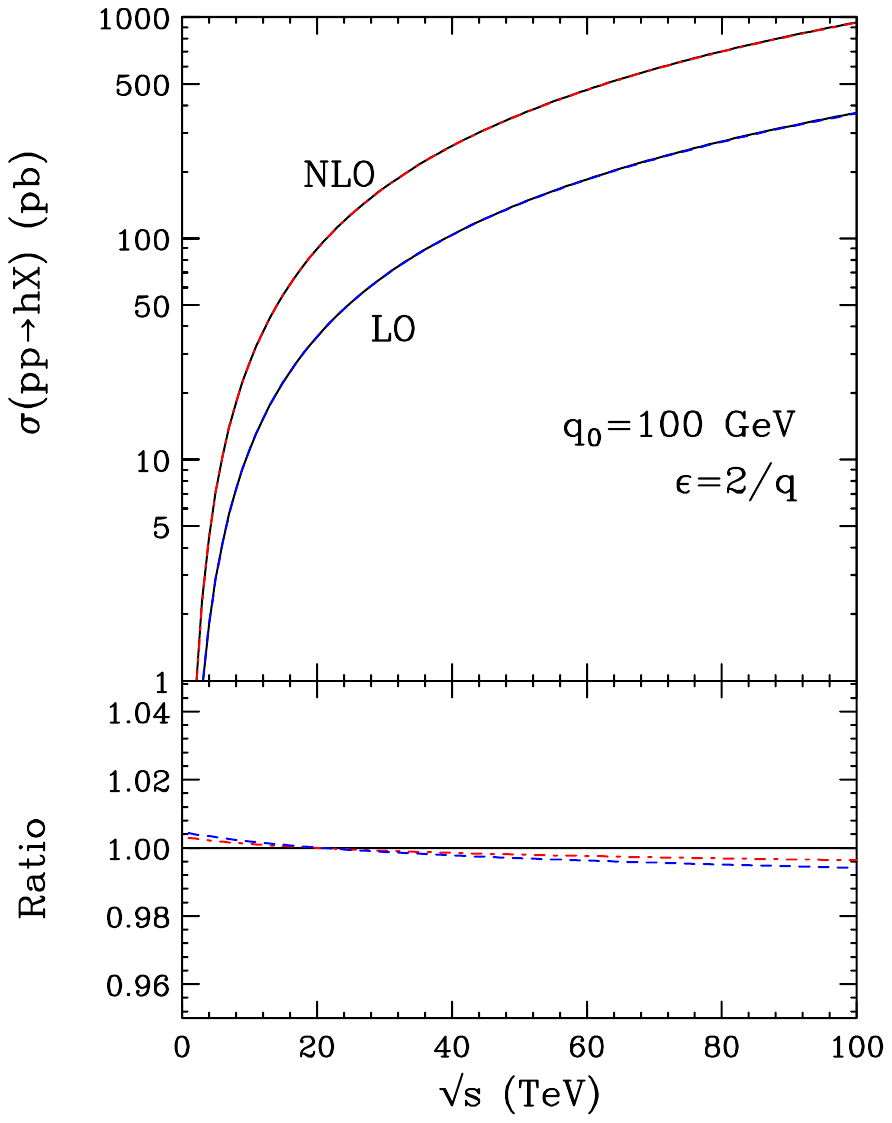}
\caption{\label{fig:hgcs}%
  Higgs cross section at $\sqrt s=1-100$ TeV (gluon fusion
  contribution only), calculated using cutoff-dependent PDFs at
  leading order (blue, dashed) and next-to-leading order (red,
  dot-dashed), compared to the corresponding results using
  cutoff-independent PDFs (solid).
   }
\end{center}
\end{figure}
%%%%%%%%%%%%%%%%%%%%%%%%%%%%%%%%%%%%%%%%%%%%%%%%%%%%%%%%%%%%%%%%%%%%%%%%%%%

Figure~\ref{fig:hgcs} shows results  (in the infinite top mass
approximation) for $pp$ collisions at $\sqrt
s=1-100$ TeV.  The PDFs, $\as$, cutoffs and factorisation scheme are as in
sect.~\ref{sec:DY}, but now the scale used in their evaluation is fixed at
$\mut=m^2_h$.  Thus the differences between the cutoff-dependent
(blue, dashed) and cutoff-independent (black, solid) LO results simply
reflect the different $x$ dependences of the corresponding gluon PDFs.
Since by construction the cutoff-dependent and -independent PDFs
coincide at the reference scale $q_0$, and the former evolve more
slowly, the corresponding LO result falls below the true (i.e.~obtained 
with cutoff-independent PDFs) value at high $\sqrt s$ (small $x$) as long 
as $q_0<m_h$.  This discrepancy is naturally much smaller for 
$q_0=100\;\mbox{GeV}\;\sim m_h$ than for $q_0=10$ GeV.

The next-to-leading order corrections to Higgs production are very
large, comparable to the leading order.  The relative differences
between the cutoff-dependent (red, dot-dashed) and true
cutoff-independent (black, solid) NLO results are reduced compared to 
leading order. Again they are naturally much smaller for $q_0=100$ GeV 
than for $q_0=10$ GeV.

\section{Conclusions\label{sec:conc}}
Our aim in this paper has been to study the extent to which the use of
PDFs to guide backward MC parton showering can be a consistent
procedure.  We have shown that it cannot be fully so if normal
cutoff-independent PDFs are used, even if the non-emission
probabilities (NEPs) currently in use in Monte Carlo event generators
(MCEGs) are corrected to account for the fact that they generate only
resolved parton emissions.  The cutoffs inherent in the resolution criteria
lead to inconsistencies that are formally power-suppressed in the
resolved region. Nevertheless, these can accumulate to have large
effects when showers evolve over a wide range of scales, and
increase with $x$.

As an alternative, formally more consistent approach, we have
considered the use of cutoff-dependent PDFs, together with
short-distance cross sections that include compensating cutoff
corrections.  We have illustrated the extent to which this
compensation works at NLO in lepton pair and Higgs boson production.

Obviously, if the use of cutoff-dependent PDFs for event generation at
hadron colliders is to be pursued, global PDF fits tailored to the
sets of cutoffs in the widely used MCEGs would need to be performed.
In principle this seems a straightforward matter of using the
cutoff-dependent PDF evolution kernels and corresponding subprocess
cutoff corrections; at leading order, this could
be a worthwhile improvement on the current practice.
Beyond leading order, however, the whole concept of guided backward
parton showering needs further clarification.

\section*{Acknowledgements}
We are grateful to Torbj\"orn Sj\"ostrand and Mike Seymour for 
valuable comments on the manuscript.
This work has been partially supported by UK STFC HEP Theory
Consolidated grant ST/T000694/1. SF thanks the CERN TH division 
for the kind hospitality during the course of this work.

\appendix
\section{PDF reconstruction with MC backward evolution\label{sec:PDFsol}}
In order to see how MC initial-state showers reconstruct PDFs, we
first need to find a solution of the evolution equations that renders 
the comparison with MC-derived results as easy as possible. To this end, 
we employ eqs.~(\ref{Oopgt}) and~(\ref{Ooplt}), and rewrite the latter 
as follows:
\beq
\Oop^\Rin(z)=\Oop_R^\Rin(z)+
\overline{\Bop}^\Rin\delta(1-z)\,,
\eeq
so that:
\beq
\Zop\left[F\right](x)=\Oop_R^\Rin\otimes_x F\,.
\label{OopRlt}
\eeq
In other words, $\Oop_R^\Rin$ is the contribution to the inner-region
evolution operator $\Oop^\Rin$ due to real (as opposed to virtual)
emissions. With eqs.~(\ref{Oopgt}) and~(\ref{OopRlt}) one obtains:
\beqn
M\Big[\Wop\left[F\right]\Big]&=&
M\Big[\Oop^\Rout\Big]M\Big[F\Big]\equiv
\Oop_N^\Rout\,F_N\,,
\\
M\Big[\Zop\left[F\right]\Big]&=&
M\Big[\Oop_R^\Rin\Big]M\Big[F\Big]\equiv
\Oop_{R,N}^\Rin\,F_N\,,
\eeqn
where by $M[g]\equiv g_N$ we have denoted the Mellin transform
of a function $g(x)$. Thus, the Mellin transform of eq.~(\ref{intevolmu}) 
reads as follows:
\beq
F_N(\mut)=\frac{\Sop(\mut)}{\Sop(\muzt)}\,F_N(\muzt)+
\int_{\muzt}^{\mut}\frac{d\kappa^2}{\kappa^2}
\frac{\Sop(\mut)}{\Sop(\kappa^2)}\,
\Big(\Oop_N^\Rout(\kappa^2)+\Oop_{R,N}^\Rin(\kappa^2)\Big)\,
F_N(\kappa^2)\,.
\label{intevolMell}
\eeq
Equation~(\ref{intevolMell}) is a
Volterra equation of the second kind\footnote{Its kernel is separable in the
two relevant variables ($\mut$ and $\kappa^2$), which leads to (at least
in a one-dimensional flavour space) a closed-form solution; this,
however, is not of particular interest here, and will not be considered.},
which is formally solved by a Neumann series:
\beq
F_N(\mut)=\sum_{k=0}^\infty F_N^{(k)}(\mut)\,,
\label{Neums}
\eeq
with:
\beqn
F_N^{(0)}(\mut)&\!\!=\!\!&\frac{\Sop(\mut)}{\Sop(\muzt)}\,F_N(\muzt)\,,
\label{Neums0}
\\
F_N^{(k)}(\mut)&\!\!=\!\!&\int_{\muzt}^{\mut}
\Bigg[\prod_{p=1}^k \frac{d\kappa_p^2}{\kappa_p^2}\,
\stepf\left(\kappa_{p+1}^2\le\kappa_p^2\le\kappa_{p-1}^2\right)
\frac{\Sop(\kappa_{p-1}^2)}{\Sop(\kappa_p^2)}
\nonumber
\\*&&\phantom{\int_{\muzt}^\infty\Bigg[a}
\times\Big(\Oop_N^\Rout(\kappa_p^2)+\Oop_{R,N}^\Rin(\kappa_p^2)\Big)
\Bigg]\frac{\Sop(\kappa_k^2)}{\Sop(\muzt)}\,
F_N(\muzt)\,,\phantom{aa}
\label{Neumsk}
\eeqn
where the matrix product in eq.~(\ref{Neumsk}) has a left-to-right 
order, i.e.~the elements corresponding to $p=1$ ($p=k$) are the leftmost
(rightmost) ones, and we have defined:
\beq
\kappa_0^2=\mut\,,\;\;\;\;\;\;\;\;
\kappa_{k+1}^2=\muzt\,.
\eeq
Equations~(\ref{Neums0}) and~(\ref{Neumsk})
can be easily transformed back to the $x$ space:
\beqn
F^{(0)}(x,\mut)&\!=\!&\frac{\Sop(\mut)}{\Sop(\muzt)}\,F(x,\muzt)\,,
\label{zNeums0}
\\
F^{(k)}(x,\mut)&\!=\!&\int_0^1
\left[\prod_{q=1}^{k+1} dz_q\right]
\delta\!\left(x-\prod_{q=1}^{k+1} z_q\right)
\label{zNeumsk}
\\*&&\!\!\!\!\times
\int_{\muzt}^{\mut}
\Bigg[\prod_{p=1}^k \frac{d\kappa_p^2}{\kappa_p^2}\,
\stepf\left(\kappa_{p+1}^2\le\kappa_p^2\le\kappa_{p-1}^2\right)
\frac{\Sop(\kappa_{p-1}^2)}{\Sop(\kappa_p^2)}
\nonumber
\\*&&\phantom{\int_{\muzt}^\infty\Bigg[a}
\times\Big(\Oop^\Rout(z_p,\kappa_p^2)+\Oop_R^\Rin(z_p,\kappa_p^2)\Big)
\Bigg]\frac{\Sop(\kappa_k^2)}{\Sop(\muzt)}\,
F(z_{k+1},\muzt)\,.\phantom{aa}
\nonumber
\eeqn
The presence of a Dirac $\delta$ in eq.~(\ref{zNeumsk}) complicates its
manipulation. We therefore introduce the $k$ independent variables:
\beq
y_i=\prod_{j=i+1}^{k+1} z_j\,,\;\;\;\;\;\;\;\;1\le i\le k\,,
\label{yidef}
\eeq
and the dummy variable
\beq
y_0\equiv x\,,
\label{yzdef}
\eeq
so that
\beq
y_i=z_{i+1}y_{i+1}\;\;\;({\rm for}~0\le i\le k-1)\,,\;\;\;\;\;\;\;\;
y_k=z_{k+1}\,.
\label{yivsxi}
\eeq
In this way, by using the identity:
\beq
1=\int_0^1\left(\prod_{i=1}^{k-1}
dy_i\,\delta\big(y_i-z_{i+1}y_{i+1}\big)\right)
dy_k\,\delta\big(y_k-z_{k+1}\big)\,,
\eeq
eq.~(\ref{zNeumsk}) becomes:
\beqn
F^{(k)}(z,\mut)&\!\!=\!\!&\int_0^1
\left[\prod_{q=1}^k \frac{dy_q}{y_q}\right]
\int_{\muzt}^{\mut}
\Bigg[\prod_{p=1}^k \frac{d\kappa_p^2}{\kappa_p^2}\,
\stepf\left(\kappa_{p+1}^2\le\kappa_p^2\le\kappa_{p-1}^2\right)
\frac{\Sop(\kappa_{p-1}^2)}{\Sop(\kappa_p^2)}
\label{zNeumsk2}
\\*&&\phantom{\int_{\muzt}^\infty\Bigg[a}
\times\left(\Oop^\Rout\left(\frac{y_{p-1}}{y_p},\kappa_p^2\right)+
\Oop_R^\Rin\left(\frac{y_{p-1}}{y_p},\kappa_p^2\right)\right)
\Bigg]\frac{\Sop(\kappa_k^2)}{\Sop(\muzt)}\,
F(y_k,\muzt)\,.\phantom{aa}
\nonumber
\eeqn
Note that from eq.~(\ref{yivsxi}):
\beq
x\equiv y_0\le y_1\le\ldots y_{k-1}\le y_k\,,
\eeq
which are automatically enforced by the requirement that the first
arguments of $\Oop^\Rout$ and $\Oop_R^\Rin$ in eq.~(\ref{zNeumsk2})
be less than one. For a given parton identity $i_0$, eq.~(\ref{zNeumsk2})
gives:
\beqn
f_{i_0}^{(1)}(x,\mut)&\!\!=\!\!&\sum_{i_1}\int_0^1
\frac{dy_1}{y_1}
\int_{\muzt}^{\mut}
\frac{d\kappa_1^2}{\kappa_1^2}\,
\frac{S_{i_0}(\mut)}{S_{i_0}(\kappa_1^2)}
\label{zNeums1}
\\*&&\phantom{aaaaa}
\times\left((\Oop^\Rout)_{i_0i_1}\!\left(\frac{x}{y_1},\kappa_1^2\right)+
(\Oop_R^\Rin)_{i_0i_1}\!\left(\frac{x}{y_1},\kappa_1^2\right)\right)
\frac{S_{i_1}(\kappa_1^2)}{S_{i_1}(\muzt)}\,
f_{i_1}(y_1,\muzt)\,,\phantom{aa}
\nonumber
\\
f_{i_0}^{(2)}(x,\mut)&\!\!=\!\!&\sum_{i_1i_2}\int_0^1
\frac{dy_1}{y_1}\frac{dy_2}{y_2}
\int_{\muzt}^{\mut}
\frac{d\kappa_1^2}{\kappa_1^2}\frac{d\kappa_2^2}{\kappa_2^2}\,
\stepf\left(\kappa_2^2\le\kappa_1^2\right)
\frac{S_{i_0}(\mut)}{S_{i_0}(\kappa_1^2)}
\label{zNeums2}
\\*&&\phantom{aaaaa}
\times\left((\Oop^\Rout)_{i_0i_1}\!\left(\frac{x}{y_1},\kappa_1^2\right)+
(\Oop_R^\Rin)_{i_0i_1}\!\left(\frac{x}{y_1},\kappa_1^2\right)\right)
\frac{S_{i_1}(\kappa_1^2)}{S_{i_1}(\kappa_2^2)}
\nonumber
\\*&&\phantom{aaaaa}
\times\left((\Oop^\Rout)_{i_1i_2}\!\left(\frac{y_1}{y_2},\kappa_2^2\right)+
(\Oop_R^\Rin)_{i_1i_2}\!\left(\frac{y_1}{y_2},\kappa_2^2\right)\right)
\frac{S_{i_2}(\kappa_2^2)}{S_{i_2}(\muzt)}\,
f_{i_2}(y_2,\muzt)\,,\phantom{aa}
\nonumber
\eeqn
and so forth.

We now assume that the parton type $i_0$, momentum fraction $x$,
and two scales $\mut$ and $\muzt$ (with $\muzt\le\mut$) are given, and
we want to compute the probability that, starting the evolution
at $(x,\mut)$, one eventually (i.e.~after an arbitrary number of
backward emissions, including none) ends up emitting at a scale 
lower than $\muzt$. Such a probability is the sum of the probabilities
$p_k$ associated with $k$ emissions, with \mbox{$0\le k\le\infty$}. As far
as $k=0$ is concerned, $p_0$ is equal to one minus the probability of
emitting at scales larger than $\muzt$, in turn equal to the non-emission
probability in $(\muzt,\mut)$ at $x$. Thus, one needs to start with a
definite choice for the latter; we begin by considering $\NEPbwR{}$ of
eq.~(\ref{BEstepR}). Hence:
\beqn
p_0&=&\NEPbwR{i_0}(x,\muzt,\mut)=
\frac{S_{i_0}(\mut)}{S_{i_0}(\muzt)}\,
\frac{f_{i_0}(x,\muzt)}{f_{i_0}(x,\mut)}\,.
\label{prob0}
\\*&=&
\frac{f_{i_0}^{(0)}(x,\mut)}{f_{i_0}(x,\mut)}\,,
\eeqn
having used eq.~(\ref{zNeums0}) in the second line.
The case $k=1$ corresponds to one emission at a scale 
\mbox{$\kappa_1^2\in(\muzt,\mut)$}, followed by one emission below
$\muzt$. As far as the probability associated with the former is
concerned, one first determines the scale at which it occurs
by solving
\beq
r=\NEPbwR{i_0}(x,\kappa_1^2,\mut)
\label{solfork1}
\eeq
for $\kappa_1^2$, given a uniform random number $r$. The solution is
discarded if $\kappa_1^2<\muzt$, which gives the correct normalisation.
Indeed, the distribution in $\log\kappa_1^2$ induced by eq.~(\ref{solfork1})
is:
\beq
\frac{\partial}{\partial\log\kappa_1^2}\,\NEPbwR{i_0}(x,\kappa_1^2,\mut)\,,
\;\;\;\;\;\;\;\;\muzt\le\kappa_1^2\le\mut\,,
\eeq
so that the total probability for such an emission is:
\beqn
\int_{\muzt}^{\mut} d\log\kappa_1^2
\frac{\partial}{\partial\log\kappa_1^2}\,\NEPbwR{i_0}(x,\kappa_1^2,\mut)&=&
\NEPbwR{i_0}(x,\mut,\mut)-\NEPbwR{i_0}(x,\muzt,\mut)
\nonumber\\*&=&
1-\NEPbwR{i_0}(x,\muzt,\mut)\,.
\eeqn
As was already reported in eq.~(\ref{dNEPrdk0}), by means of an explicit 
computation we obtain:
\beq
\frac{\partial}{\partial\log\kappa_1^2}\,\NEPbwR{i_0}(x,\kappa_1^2,\mut)=
\frac{1}{f_{i_0}(x,\mut)}\frac{S_{i_0}(\mut)}{S_{i_0}(\kappa_1^2)}
\Big[\big(\Wop\left[F\right]\!\big)_{i_0}(x,\kappa_1^2)+
\big(\Zop\left[F\right]\!\big)_{i_0}(x,\kappa_1^2)\Big].
\label{dNEPrdk}
\eeq
Having computed the probability density for an emission at $\kappa_1^2>\muzt$,
we need to multiply it by the probability of the next emission occurring
below $\muzt$. This is functionally the same quantity as was computed 
for the $k=0$ case in eq.~(\ref{prob0}); presently, we must use that
result by replacing $\mut\to\kappa_1^2$, and $z$ and $i_0$ with the
momentum fraction $y_1$ and parton type $i_1$ that have resulted from 
the branching at $\kappa_1^2$, which may have changed w.r.t.~the 
original $z$ and $i_0$. In order to determine these, and the probabilities
associated with their choices, we introduce the functions:
\beq
{\cal P}_{ij}(y;x,\kappa^2)=
\int_y^1 \frac{d\omega}{\omega}
\left[(\Oop^\Rout)_{ij}\left(\frac{x}{\omega},\kappa^2\right)+
(\Oop_R^\Rin)_{ij}\left(\frac{x}{\omega},\kappa^2\right)\right]
f_j(\omega,\kappa^2)\,,
\label{Pijdef}
\eeq
with $y\ge x$; these are such that:
\beq
\sum_j{\cal P}_{ij}(x;x,\kappa^2)=
\big(\Wop\left[F\right]\!\big)_i(x,\kappa^2)+
\big(\Zop\left[F\right]\!\big)_i(x,\kappa^2)\,.
\label{Pijnorm}
\eeq
We first define $i_1$ to be the smallest index that fulfills the
following inequality:
\beq
r\le\frac{\sum_j^{j\le i_1}{\cal P}_{i_0j}(x;x,\kappa_1^2)}
{\sum_j{\cal P}_{i_0j}(x;x,\kappa_1^2)}\,,
\eeq
with $r$ a uniform random number; in this way, the probability of
obtaining a given $i_1$ is equal to:
\beq
\frac{{\cal P}_{i_0i_1}(x;x,\kappa_1^2)}
{\big(\Wop\left[F\right]\!\big)_{i_0}(x,\kappa_1^2)+
\big(\Zop\left[F\right]\!\big)_{i_0}(x,\kappa_1^2)}\,,
\label{probi1}
\eeq
owing to eq.~(\ref{Pijnorm}). Next, we determine $y_1$ by solving
for it the equation:
\beq
1-r=\frac{{\cal P}_{i_0i_1}(y_1;x,\kappa_1^2)}
{{\cal P}_{i_0i_1}(x;x,\kappa_1^2)}\,,
\label{proby10}
\eeq
with $r$ a uniform random number. Thus, the probability distribution
associated with $y_1$ is:
\beqn
&&-\frac{\partial}{\partial y_1}\,
\frac{{\cal P}_{i_0i_1}(y_1;x,\kappa_1^2)}
{{\cal P}_{i_0i_1}(x;x,\kappa_1^2)}=
\label{proby1}
\\*&&\phantom{aaaaaa}
\frac{1}{{\cal P}_{i_0i_1}(x;x,\kappa_1^2)}
\frac{1}{y_1}
\left[(\Oop^\Rout)_{i_0i_1}\left(\frac{x}{y_1},\kappa_1^2\right)+
(\Oop_R^\Rin)_{i_0i_1}\left(\frac{x}{y_1},\kappa_1^2\right)\right]
f_{i_1}(y_1,\kappa_1^2)\,.\phantom{aa}
\nonumber
\eeqn
We finally obtain the sought probability by multiplying the
results of eqs.~(\ref{dNEPrdk}), (\ref{probi1}), (\ref{proby1}),
and~(\ref{prob0}) (with $\mut\to\kappa_1^2$, $i_0\to i_1$,
and $x\to y_1$ in the latter), by integrating over all possible
intermediate scales $\kappa_1^2$ and momentum fractions $y_1$, and 
by summing over all possible parton types $i_1$:
\beqn
p_1&=&\sum_{i_1}\int_0^1 dy_1\int_{\log\muzt}^{\log\mut} d\log\kappa_1^2
\nonumber\\*&&\phantom{aaa}\times
\frac{1}{f_{i_0}(x,\mut)}\frac{S_{i_0}(\mut)}{S_{i_0}(\kappa_1^2)}
\Big[\big(\Wop\left[F\right]\!\big)_{i_0}(x,\kappa_1^2)+
\big(\Zop\left[F\right]\!\big)_{i_0}(x,\kappa_1^2)\Big]
\nonumber\\*&&\phantom{aaa}\times
\frac{{\cal P}_{i_0i_1}(x;x,\kappa_1^2)}
{\big(\Wop\left[F\right]\!\big)_{i_0}(x,\kappa_1^2)+
\big(\Zop\left[F\right]\!\big)_{i_0}(x,\kappa_1^2)}
\nonumber\\*&&\phantom{aaa}\times
\frac{1}{{\cal P}_{i_0i_1}(x;x,\kappa_1^2)}
\frac{1}{y_1}
\left[(\Oop^\Rout)_{i_0i_1}\left(\frac{x}{y_1},\kappa_1^2\right)+
(\Oop_R^\Rin)_{i_0i_1}\left(\frac{x}{y_1},\kappa_1^2\right)\right]
f_{i_1}(y_1,\kappa_1^2)
\nonumber\\*&&\phantom{aaa}\times
\frac{S_{i_1}(\kappa_1^2)}{S_{i_1}(\muzt)}\,
\frac{f_{i_1}(y_1,\muzt)}{f_{i_1}(y_1,\kappa_1^2)}\,.
\eeqn
Therefore, from eq.~(\ref{zNeums1}):
\beq
p_1=\frac{f_{i_0}^{(1)}(x,\mut)}{f_{i_0}(x,\mut)}\,.
\eeq
This procedure can manifestly be iterated, to obtain:
\beq
p_k=\frac{f_{i_0}^{(k)}(x,\mut)}{f_{i_0}(x,\mut)}
\;\;\;\;\Longrightarrow\;\;\;\;
\sum_{k=0}^\infty p_k=1\,,
\label{NEPrunit}
\eeq
having exploited the inverse Mellin transform of eq.~(\ref{Neums}).

Equation~(\ref{NEPrunit}) shows that an evolution generated
by means of $\NEPbwR{}$ of eq.~(\ref{BEstepR}) and of the functions of
eq.~(\ref{Pijdef}) for the backward steps in the scale and $x$
spaces, respectively, allows one to recover the PDF used to guide
the evolution. However, it should be clear that this conclusion is affected
by a number of fallacies, that have to do with the NEP possibly
being non-monotonic and not bounded from above by one, as well as the
probabilities of eq.~(\ref{probi1}) possibly being negative. Both aspects 
have ultimately to do with the fact that $\NEPbwR{}$ of eq.~(\ref{BEstepR}) 
also accounts for non-resolvable contributions;
as such, it is consistent that it be in agreement (although only
formally) in the sense of eq.~(\ref{NEPrunit}) with the PDF, whose
form is determined by both resolved and non-resolved contributions.
Moreover, note that eq.~(\ref{Pijdef}) is {\em not} what is current 
employed in practical MC implementations, which rather corresponds to that 
form with the $\Oop^\Rout$ contribution removed (see e.g.~eq.~(\ref{dPjodx})): 
the proof above shows that, by doing so, one does not recover the PDF after 
the evolution.

One can repeat this procedure by adopting the true NEP of either 
eq.~(\ref{BnepJ}) or eq.~(\ref{BnepJwZ}) (the two coincide).
The analogue of eq.~(\ref{dNEPrdk}) is (see eq.~(\ref{dNEPdl0})):
\beq
\frac{\partial}{\partial\log\kappa_1^2}\,\NEPbw{i_0}(x,\kappa_1^2,\mut)=
\frac{1}{f_{i_0}(x,\mut)}\frac{S_{i_0}(\mut)}{S_{i_0}(\kappa_1^2)}\,
\big(\Zop\left[F\right]\!\big)_{i_0}(x,\kappa_1^2)\,.
\label{dNEPdl}
\eeq
Because of this result, the analogues of the functions ${\cal P}_{ij}$
to be employed in this case are obtained from those in eq.~(\ref{Pijdef}) 
by removing the $\Oop^\Rout$ contribution there. By doing so, one
arrives at the analogue of eq.~(\ref{NEPrunit}), which reads:
\beq
p_k=\frac{1}{f_{i_0}(x,\mut)}
\left(\left.f_{i_0}^{(k)}(x,\mut)\right|_{\Oop^\Rout\to 0}\right)
\;\;\;\;\Longrightarrow\;\;\;\;
\sum_{k=0}^\infty p_k\ne 1\,.
\label{NEPunit}
\eeq
This result need not be surprising: $\NEPbwi$ correctly accounts for 
resolved emissions only, while as was already said an actual PDF includes 
non-resolved contributions. It may appear that the PDF could be recovered
in the context of this evolution by including a branching-by-branching
correction factor equal to:
\beq
\left[(\Oop^\Rout)_{i_{p-1}i_p}\left(\frac{y_{p-1}}{y_p},\kappa_p^2\right)+
(\Oop_R^\Rin)_{i_{p-1}i_p}\left(\frac{y_{p-1}}{y_p},\kappa_p^2\right)\right]
\Bigg/
(\Oop_R^\Rin)_{i_{p-1}i_p}\left(\frac{y_{p-1}}{y_p},\kappa_p^2\right)\,,
\label{Orat}
\eeq
as well as correcting by means of the ratio $\NEPbwR{i_0}/\NEPbw{i_0}$ 
the zero-emission contribution. Unfortunately, eq.~(\ref{Orat}) does not
work: $\Oop^\Rout$ and $\Oop_R^\Rin$ have non-overlapping supports in the
$x$ space, and here $y_p$ has been generated by using only the latter
operator; it follows that the ratio of eq.~(\ref{Orat}) is in practice 
always equal to one.

Finally, when adopting $\NEPbwE{}$~(\ref{BEstepE}) for the NEP, the analogue 
of eq.~(\ref{dNEPrdk}) reads as follows (see eq.~(\ref{dNEPEdl0})):
\beqn
&&\frac{\partial}{\partial\log\kappa_1^2}\,\NEPbwE{i_0}(x,\kappa_1^2,\mut)=
\label{dNEPEdl}
\\*&&\phantom{aaaaaa}
\frac{1}{f_{i_0}(x,\mut)}\frac{S_{i_0}(\mut)}{S_{i_0}(\kappa_1^2)}\,
\big(\Zop\left[F\right]\!\big)_{i_0}(x,\kappa_1^2)\,
\exp\!\left[\int_{\kappa_1^2}^{\mut}\frac{d\kappa^2}{\kappa^2}
\frac{1}{f_{i_0}(x,\kappa^2)}\,
\big(\Wop\left[F\right]\!\big)_{i_0}(x,\kappa^2)\right],
\nonumber
\eeqn
having employed eq.~(\ref{intevolexpmu}). The similarity of this
result with that of eq.~(\ref{dNEPdl}) suggests that also in this 
case one can obtain a PDF that stems from keeping only resolved emissions
by including a branching-by-branching correction factor equal to:
\beq
\exp\left[\int_{\kappa_p^2}^{\kappa_{p-1}^2}\frac{d\kappa^2}{\kappa^2}
\frac{1}{f_{i_{p-1}}(y_{p-1},\kappa^2)}\,
\big(\Wop\left[F\right]\!\big)_{i_{p-1}}(y_{p-1},\kappa^2)\right].
\eeq
Since $\Wop[F]$ has no definite sign, this factor can be larger or smaller
than one. This is connected with the fact that the $\ord(\as^2)$ coefficient
in eq.~(\ref{diffNEPE3}) has no definite sign, and its cumulative effect
over successive backward branchings is therefore typically smaller than
naive coupling-constant power counting would suggest, a point borne out
in practice by the results shown in fig.~\ref{fig:NEPgu}.

\section{An academic model\label{sec:purevirt}}
A different perspective on the three forms of NEP considered in this
paper can be obtained in the context of an academic model, defined
so that the only branchings are of virtual origin. This can be
achieved by setting:
\beq
\Aop=\Cop=0\,.
\label{ACvirt}
\eeq
Before proceeding, we stress that eq.~(\ref{ACvirt}) defines the 
virtual contribution in an unique manner only because one understands 
eq.~(\ref{Oopform}) so that, from eq.~(\ref{bBop})
\beq
\overline{\Bop}=\Bop\,.
\label{virt1}
\eeq
A different model which still sets the real splitting kernels
equal to zero can be defined as follows:
\beq
\widetilde{\Aop}=\Cop=0
\label{Atilvirt}
\eeq
which understands the form of eq.~(\ref{Oopform2}), so that
from eq.~(\ref{bBop2}):
\beq
\overline{\Bop}=\widetilde{\Bop}\,.
\label{virt2}
\eeq
Thus: $\Aop=0$ implies eq.~(\ref{virt1}), while $\widetilde{\Aop}=0$
implies eq.~(\ref{virt2}), with $\Bop$ and $\widetilde{\Bop}$ still
related to one another by eq.~(\ref{ABopid}), with $\widetilde{\Aop}\ne 0$
there. For example, in the case of the gluon at the LO in QCD:
\beqn
&&A_g(z)=0\;\;\;\;\Longrightarrow\;\;\;\;
\tpioas\,\overline{B}_g=-\frac{\CA+4\TF\NF}{6}\,,
\label{virt1g}
\\
&&\widetilde{A}_g(z)=0\;\;\;\;\Longrightarrow\;\;\;\;
\tpioas\,\overline{B}_g=\gamma(g)\,.
\label{virt2g}
\eeqn
Having clarified this point, we can solve the PDF evolution equations
and consider the MC-generated backward evolution for the model of
eq.~(\ref{ACvirt}). We do so by employing the parameter $\lambda$
introduced in eq.~(\ref{lamrepl}). We can solve
directly the PDF evolution equations, and obtain:
\beq
F(\mut)=\frac{\Sop_{\lambda=0}(\mut)}{\Sop_{\lambda=0}(\muzt)}\,F(\muzt)\,.
\label{solvirt}
\eeq
With eq.~(\ref{intevolmu}) we have instead
\beq
F(\mut)=\frac{\Sop(\mut)}{\Sop(\muzt)}\,F(\muzt)+
\lambda\int_{\muzt}^{\mut}\frac{d\kappa^2}{\kappa^2}
\frac{\Sop(\mut)}{\Sop(\kappa^2)}\,
\overline{\Bop}^\Rout(\kappa^2) F(\kappa^2)\,.
\label{intevolmuappvirt}
\eeq
These two solutions coincide (as they should), since by using 
eq.~(\ref{solvirt}) in the second term on the r.h.s.~of
eq.~(\ref{intevolmuappvirt}) one obtains:
\beq
\lambda\int_{\muzt}^{\mut}\frac{d\kappa^2}{\kappa^2}
\frac{\Sop(\mut)}{\Sop(\kappa^2)}\,
\overline{\Bop}^\Rout(\kappa^2) F(\kappa^2)=
\frac{\Sop_{\lambda=0}(\mut)}{\Sop_{\lambda=0}(\muzt)}\,F(\muzt)-
\frac{\Sop(\mut)}{\Sop(\muzt)}\,F(\muzt)\,,
\eeq
which shows explicitly, in this simple case, the cancellation of
dependence on $\lambda$ on the r.h.s.~of eq.~(\ref{intevolmu}).

We now consider an MC backward evolution. We start by noting that:
\beq
\NEPbwi=1\,,\;\;\;\;\;\;\;\;
\NEPbwiE=1\,,\;\;\;\;\;\;\;\;
\NEPbwiR=\frac{S_{i,\lambda=0}(\muzt)}{S_{i,\lambda=0}(\mut)}\,
\frac{S_i(\mut)}{S_i(\muzt)}\,.
\label{NEPsvirt}
\eeq
The leftmost result in eq.~(\ref{NEPsvirt}) is what we expect
in view of the physical interpretation of $\NEPbwi$: only strictly
resolved emissions may contribute to it, and in this model no
resolved emissions can be generated -- thus, the NEP must be
equal to one. The middle result in eq.~(\ref{NEPsvirt}) shows
that this model is too simple to allow one to distinguish the
behaviour of $\NEPbwiE$ from that of $\NEPbwi$: the spurious 
terms of $\Wop[F]$ origin potentially present in the former case
according to eq.~(\ref{dNEPEdl0}) are all identically equal to zero,
being proportional to $\Zop[F]=0$. Therefore, also in this case the
NEP is equal to one. Finally, the rightmost result in eq.~(\ref{NEPsvirt}) 
shows that for any $\lambda\ne 0$ the NEP is not equal to one since
it receives, independently from one another, both resolved and 
unresolved contributions, and the latter are different from zero
owing to the virtual contribution to $\Wop[F]$ (see eq.~(\ref{WJop2})).
Having said that, we point out that, in view of eq.~(\ref{solvirt}),
a sound probabilistic interpretation of $\NEPbwiR$ requires that:
\beq
S_i(\mut)<S_i(\muzt)\;\;\;\;\Longleftrightarrow\;\;\;\;
\overline{B}_i^\Rin\equiv\overline{B}_i-\lambda\overline{B}_i^\Rout<0\,,
\eeq
for any $\mut>\muzt$. Therefore, in the simplest scenario $\lambda=0$,
this happens for the gluon (see eq.~(\ref{virt1g})) in the context
of the model of eq.~(\ref{ACvirt}) we work with; however, this would
{\em not} happen had we chosen the different model of eq.~(\ref{Atilvirt})
(see eq.~(\ref{virt2g})). This simple example confirms a general fact that 
has already been inferred before, namely that the interpretation of 
$\NEPbwiR$ as a NEP might become problematic.

If we want to obtain the Neumann summands of eq.~(\ref{zNeumsk2})
relevant to the model of eq.~(\ref{ACvirt}) we need to use the
fact that:
\beq
(\Oop^\Rout)_{ij}\left(z,\kappa^2\right)=\lambda\delta_{ij}
\overline{B}_i^\Rout(\kappa^2)\delta(1-z)\,,
\;\;\;\;\;\;\;\;
(\Oop_R^\Rin)_{ij}\left(z,\kappa^2\right)=0\,.
\label{OopACvirt}
\eeq
The absence of off-diagonal terms leads to an immediate dramatic
simplification of eq.~(\ref{zNeumsk2}), which becomes:
\beq
f_{i_0}^{(k)}(x,\mut)=
\frac{S_{i_0}(\mut)}{S_{i_0}(\muzt)}\,f_{i_0}(x,\muzt)\,
\frac{\lambda^k}{k!}\int_{\muzt}^{\mut}
\frac{d\kappa^2}{\kappa^2}\overline{B}_{i_0}^\Rout(\kappa^2)\,,
\label{Neusvirt}
\eeq
a result that is also valid for $k=0$. By summing over $k$ one finds 
again the solution of eq.~(\ref{solvirt}), as one must by construction.
Moreover, as we have previous learned, eq.~(\ref{Neusvirt}) can be seen
as the MC contribution to the PDFs due to showers that feature $k$
emissions. However, for this to be true in the model defined by
eq.~(\ref{ACvirt}), one would have to have chosen $\NEPbwiR$ as the
NEP, since that is the only nontrivial NEP in this context
(see eq.~(\ref{NEPsvirt})). Therefore, this simple example confirms 
the previous general findings, namely that while $\NEPbwiR$ is not the 
correct non-emission probability, it nevertheless formally allows one to 
recover the PDF given in input. Conversely, if $\NEPbwi$ (or $\NEPbwiE$) 
had been adopted, both matrix elements in the analogue of eq.~(\ref{OopACvirt}) 
would be equal to zero, leading to a Neumann series whose terms would be
all equal to zero bar the first. The latter then coincides with 
the reconstructed PDF, and reads as follows:
\beq
f_{i_0}^{(0)}(x,\mut)=
\frac{S_{i_0}(\mut)}{S_{i_0}(\muzt)}\,f_{i_0}(x,\muzt)\,.
\eeq
This is in general different from the exact solution of
eq.~(\ref{solvirt}). Clearly, the model of eq.~(\ref{ACvirt}) is
maximally perverse, since all emissions are unresolved; it is therefore
not particularly surprising that evolutions based on NEPs that
can only account for resolved emissions fail.

\phantomsection
\addcontentsline{toc}{section}{References}
\bibliographystyle{JHEP}
\bibliography{pdfc}

\end{document}